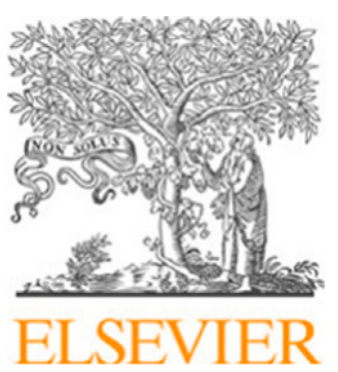

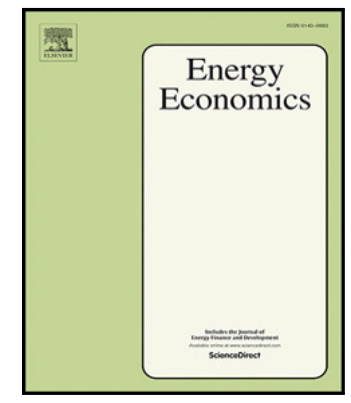

# A pricing mechanism to jointly mitigate market power and environmental externalities in electricity markets[☆]

Lamia Varawala [a,*], Mohammad Reza Hesamzadeh [a], György Dán [a], Derek Bunn [b], Juan Rosellón [c,d,e,f]

[a] KTH Royal Institute of Technology, Brinellvägen 8, 114 28 Stockholm, Sweden
[b] London Business School, Regent's Park, London NW1 4SA, United Kingdom
[c] Centro de Investigación y Docencia Económicas, Carr. México-Toluca 3655, Santa Fe, Altavista, Álvaro Obregón, 01210 Ciudad de México, CDMX, Mexico
[d] DIW Berlin, Mohrenstraße 58, 10117 Berlin, Germany
[e] Baker Institute for Public Policy, 6100 Main Street, Baker Hall MS-40, Suite 120, Houston, TX 77005, United States
[f] Universidad de Barcelona, Tinent Coronel Valenzuela, 1-11, 08034, Barcelona, Spain

ARTICLE INFO

ABSTRACT

The electricity industry has been one of the first to face technological changes motivated by sustainability concerns. Whilst efficiency aspects of market design have tended to focus upon market power concerns, the new policy challenges emphasise sustainability. We argue that market designs need to develop remedies for market conduct integrated with regard to environmental externalities. Accordingly, we develop an incentive-based market clearing mechanism using a power network representation with a distinctive feature of incomplete information regarding generation costs. The shortcomings of price caps to mitigate market power, in this context, are overcome with the proposed mechanism.

## 1. Introduction

The emergence of competitive markets for electricity has presented a particularly challenging area for research and policy related to their efficient market arrangements. Whilst specific regulatory approaches vary globally, the market designs for power must seek to adapt to the technical complexity of electricity supply systems with both spatial and temporal constraints in market clearing as well as seeking to provide financially complete microstructures for the hedging and risk management by participants. Furthermore, this must be achieved whilst recognising that regulatory surveillance will generally be inevitable to deal with possible scarcity pricing and excess producer rents due to imperfect competition. Moreover, the social and political scrutiny to which electricity becomes subject for prices and reliability sharpens the importance of avoiding mistakes in market designs. Thus, electricity provided an early application for performance-based regulation, e.g., the Incremental Surplus Subsidy of Sappington and Sibley (1988), and for auction theory, as reviewed by Wilson (2002), with the supply function equilibria of Klemperer and Meyer (1989) being applied by Green and Newbery (1992) to the analysis of inadequate competition in the new electricity market in England and Wales. Thereafter followed a substantial stream of gaming, conduct and incentive research, e.g., by Borenstein et al. (2002), and Joskow and Tirole (2000), Fabra et al. (2006), among many. Whilst the body of knowledge on the efficient design of wholesale electricity has become substantial, many of the issues which are emerging, e.g., regarding global warming, relate to the externalities of electricity, and this opens up a wider perspective on market efficiency and new questions regarding the market arrangements.

Remarkably, more electricity regulators are now being directed to include sustainability objectives in their actions and incentives to influence participant conduct. For example, in the EU, the latest policy directive, the so called *Clean Energy for All Package*[1] requires market mechanisms to evolve to meet decarbonisation targets, whilst in Britain, the electricity regulator, Ofgem has explicit policy targets to align with the Net Zero 2050 target.[2] These wider regulatory objectives require new electricity market clearing principles to emerge.

☆ Funding: This work was supported by the Digitalization of Swedish Electrical Engineering project at KTH [grant number E-2016-0277].

* Corresponding author.
*E-mail addresses:* varawala@kth.se (L. Varawala), mrhesamzadeh@ee.kth.se (M.R. Hesamzadeh), gyuri@kth.se (G. Dán), dbunn@london.edu (D. Bunn), juan.rosellon@cide.edu (J. Rosellón).

[1] Clean energy for All Europeans Package, https://ec.europa.eu/energy/topics/energy-strategy/clean-energy-all-europeans_en, Accessed: 2021–08–11.

[2] Our Strategy and Priorities, https://www.ofgem.gov.uk/about-us/our-role-and-responsibilities Accessed: 2021–08–11.

Thus, in liberalised electricity markets, an Independent System Operator (ISO) usually has the regulated obligation to clear the market according to a least cost dispatch algorithm based on the supply and demand information provided by the market participants. The ISO therefore ensures the technical performance of the power system in real time as well as facilitating competition between wholesale producers and consumers. Externalities are not directly observed in the market and are therefore not generally part of the ISO's regulated objectives in clearing the market. Thus, they are often considered separately, e.g., with carbon taxes (Maryniak et al., 2019) on producers or via separate markets for carbon allowances (Limpaitoon et al., 2011). However, it is not a matter of principle that externalities should be excluded from the ISO's consideration. System operators with substantial reservoir and hydro resources, for example, often have to dispatch against multiple ecological, environmental and agricultural constraints (Rand, 2018). Similarly, they have sometimes been directed to give priority dispatch to the wind and solar facilities.[3] Furthermore, pricing formulas have variously included considerations of security (through loss-of-load, e.g., by Cañizares et al., 2001) and there is increasing interest in developing stochastic clearing because of the increase in intermittency (two stage recourse clearing, e.g., by Tang et al., 2019). Consistent with this, it is therefore a research question whether it is feasible to include externalities directly and more explicitly in an ex ante market pricing formulation. We find it is and we develop an analytical solution, capable of computational scalability in a nodal market clearing context that is incentive compatible with social welfare maximisation, widely interpreted to include multiple externalities.

Furthermore, electricity markets often manifest concerns about the abuse of market power by producers. Surprisingly, whilst the ex post economic analysis of market power has been extensive and most commonly based around the theory of loss in social welfare (*deadweight loss*) as summarised by Prabhakar Karthikeyan et al. (2013), the design of ISO dispatching rules that would be welfare maximising, ex ante, and thereby incentivise nonabusive behaviour, have received little theoretical attention. Yet, again, this is not due to any fundamental principle that participant misconduct should only be constrained after evidence; indeed the principle of modifying the market settlement prices ex ante is well established in the real-time balancing mechanisms. In order to disincentivise market participants from going out of balance in real-time, the ISOs have been directed in several jurisdictions to introduce penalties into the imbalance prices (Shinde et al., 2021). We therefore consider if, and how, a theoretical extension of this principle of modifying the market price mechanisms, ex ante, could be extended to the wholesale markets in order to disincentivise the exercise of market power and thereby reduce the need for ad-hoc remedies. We develop an implementable solution to this proposition alongside the inclusion of externalities.

It is not just a matter of convenience to include both externalities and market power mitigation together in the pricing mechanism. To the extent that the exercise of market power can lead to the use of more expensive marginal generating units, which, in turn, may be more carbon intensive, pricing the externality costs together with the exercise of market power can create a stronger disincentive. Whilst there are precedents for an incentive compatible approach in principle, we present a radically different market clearing process to the current theory for electricity spot markets, which potentially serves to guide regulatory approaches in practice. As such, we consider the research contribution to be novel in its methodology, computability and in raising the consideration of ex ante market arrangements that would preclude, to perhaps a substantial extent, both the design of separate externality mechanisms and the ex post remedies for market power.

[3] EWEA position paper on priority dispatch of wind power, https://www.ewea.org/fileadmin/files/library/publications/position-papers/EWEA_position_on_priority_dispatch.pdf, Accessed: 2021–08–11.

The structure of the article is as follows. Following a brief review of some background research, we formulate our basic electricity spot market model in Section 3, in which Section 3.1 presents socially optimal market clearing with the impact of pollution as the environmental externality. We use the term pollution to include a wide range of measurable adverse effects, such as particulates and nitrogen oxides as well as greenhouse gas emissions. In Section 3.2, we show how pollution causes competitive market clearing to deviate from the social optimum whereas in Section 3.3, we focus on deviations from the optimum due to producers exercising market power. In Section 4, we review solutions to overcome these deviations. First, in Section 4.1, we discuss price caps as a commonly used remedy to mitigate market power and highlight their inability to address externalities as well as other shortcomings. Then, in Section 4.2, we propose an incentive compatible pricing mechanism which remedies both of these issues and highlight its properties. In Section 5, we present an example to analyse the usefulness the proposed mechanism and in Section 6, we conclude the article. Finally, in the appendices, we discuss potential extensions.

## 2. Research background

For reasons of resource adequacy and perhaps political influence, government policies have often been tolerant of market concentration in electricity generation, but at the cost of dominant producers and the consequent need for market power surveillance and mitigation. Hence, market power analysis has been one of the main themes in numerous studies, e.g., by Kumar David and Wen (2001), Green and Newbery (1992), Joskow (1997), Wolfram (1999), Guo and Shmaya (2019). Reducing market concentration is fundamental to market power mitigation (Green and Newbery, 1992; Green, 1996a; Borenstein et al., 1995) and this may require interventions by the competition authorities (Brennan and Melanie, 1998) because the electricity generating sector does not generally facilitate ease of entry (Acutt and Elliott, 1999). The divestitures of assets (Day and Bunn, 2001) is an ad-hoc remedy, however, often being awkward, slow, and especially prone to legal challenges. A more expedient remedy is the use of price caps (Vogelsang and Finsinger, 1979) as these are generally easier to apply by the regulatory authorities. As a consequence, price caps have been widely applied, as needed, around the world (Arocena and Waddams Price, 2002; Green, 1996b). Whilst they are very effective by construction – the prices do not exceed the caps – they are nevertheless often quite inefficient to the extent that they are generally set too high (Hogan, 2013).

In general, price regulation can apply to both the level and the structure of prices. Price level regulation seeks long-run distribution of rents and risks between consumers and the firm. It aims to achieve allocative, productive, and distributive efficiencies. Caps on price level are typically combined with cost-plus regulation to deliver a cost-based initial price limit that stays fixed over a regulatory lag, usually only altered by inflation and efficiency factors (RPI-X regulation). Regulation of price structure implies short-run allocation of costs and benefits among distinct categories of consumers or markets. It fosters convergence to Ramsey-Boiteux equilibrium through a cap set over an index of prices, typically calculated as the weighted sum of distinct consumer prices.

Price structure regulation is used by Vogelsang (2001) and Hogan and Schmalensee (2010) to regulate market power in electricity transmission and resolve congestion, in the short run, as well as capital costs and investment matters, in the long run. In a two-part tariff-cap model (with usage and capacity fees), the usage fee relies on nodal prices and reflects congestion, whereas the capacity charge recuperates long-term capital costs. Both Vogelsang (2001) and Hogan and Schmalensee (2010) rely on assumptions of perfect competition. When there is market power in generation, prices would not reflect generation marginal costs (Joskow and Tirole, 2005) because generators in constrained regions will tend to withdraw capacity to increase prices, which would overestimate cost-saving gains from investments in transmission. Furthermore, dominance in the market for financial transmission rights (FTRs) provides incentives to curtail output (demand) to make FTRs more valuable.

Whilst incentive-based methods are favoured in some jurisdictions, they tend to have been more applicable to retail price controls than for moderating the market power in wholesale prices. They seek to align the firms' profit maximisation with the social welfare maximisation by introducing penalties or rewards into its actions. However, Loeb and Magat (1979) proposed a mechanism that provides the consumer surplus to a monopolist firm. This radical theory relieves the regulator's informational burden about the firms' cost functions but requires knowledge of the consumers' utility function. This idea resurfaced with Gans and King (2000) who showed that, in the case of electricity wholesale markets, consumer utilities can be inferred from the locational spot prices. Hesamzadeh et al. (2018) applied the idea to power system transmission capacity investment by developing an incentive model for wholesale prices to integrate the original approach of Loeb and Magat (1979) with the price cap approach of Hogan and Schmalensee (2010). This combination of price caps and subsidies promotes immediate convergence to Ramsey-Boiteux equilibrium, disregarding the need to use price weights. It also achieves global equilibria, as opposed to piecewise equilibria as done by Hogan and Schmalensee (2010).

Going further, in order to include both externalities and market power mitigation, Kim and Chang (1993) present a general incentive compatible mechanism considering pollution as an externality by introducing a continuous pollution abatement parameter. The amount of pollution, and hence the externality, is decreasing with this parameter, but the cost to the firm is increasing. However, there are two reasons why this does not apply well to electricity markets. First, a single firm in the power system may employ a variety of generation technologies and these generation technologies cannot be represented by a continuous parameter. Second, in the power system, the short-run marginal cost of generation by a technology and the amount of pollution caused are typically positively correlated. For example, renewable energy plants cause the least amount of pollution and also have the lowest short-run marginal generation costs. To account for this, our work considers a set of discrete generation units, each with its individual pollution and cost function, without making any restrictions on correlations. For the firm (producer, in our case), we consider an aggregate of these units. Also, we consider the transmission system with nodal (or zonal) market pricing, allowing us to explicitly model the dependence of the price at one node on the generation at another. Another motivation behind considering multiple units is to more accurately represent aggregators who facilitate the market engagements of the increasing numbers of small-scale generators at the local level in the larger electricity market. We therefore contribute a novel integrative incentive based pricing theory for wholesale electricity to restrain market power and include externalities from an ex ante welfare maximising objective. Furthermore, we extend the theory to a computable solution for realistic systems involving network and unit operational details.

## 3. Electricity spot market equilibria

In what follows, we present briefly the basic principles of externalities, marginal cost pricing, from uniform auctions with market power, in a special way that is conducive to deriving the new results which we present subsequently. We consider a power system with a set $\mathcal{N}$ of nodes. Here, the term *node* is used in general and may refer to a node or a zone based on whether the market pricing system is nodal or zonal respectively. Nodes are connected to one another by a set of transmission lines $\mathcal{L}$.

We consider an electricity spot market with a set $\mathcal{I}$ of producers forming an oligopoly. Here, the term *producer* is used in general to represent conventional companies in control of larger generation units connected to the transmission system and aggregators who represent small-scale producers, typically in the local distribution networks and increasingly sited amongst consumers. A producer may be present at multiple nodes and may be using a variety of generating facilities (units) $\mathcal{J}$. We use indices $n, n', n'' \in \mathcal{N}$ to refer to individual nodes, index $l \in \mathcal{L}$ to refer to individual lines, indices $i, i' \in \mathcal{I}$ to refer to individual producers, and indices $j, j' \in \mathcal{J}$ to refer to individual units. For simplicity of notation, we use natural numbers as indices everywhere in this article, e.g., $n, n', n'' \in \mathbb{N}_0, 1 \leq n, n', n'' \leq |\mathcal{N}|$.

We denote by $q^i_{nj}$ and $k^i_{nj} \geq 0$ the power generation level and the available generation capacity, respectively, at node $n$ for producer $i$ and unit $j$, and enforce the generation capacity constraints

$$0 \leq q^i_{nj} \leq k^i_{nj}. \tag{1}$$

We denote by $C^i_{nj}\left(q^i_{nj}\right)$ the generation cost function, and we assume $C^i_{nj}(0) = 0$, and that it is differentiable, increasing and convex in $q^i_{nj}$. Throughout the article we adopt the convention that omitting an index for a quantity represents summation over that index, e.g., $q^i_n = \sum_j q^i_{nj}$. Also, to represent vectors compactly, instead of the set builder notation we use square brackets with the running index as a subscript or a superscript on the square brackets, e.g., $\left[q^i_{nj}\right]^i_j = \left(q^i_{nj} | i \in \mathcal{I}, j \in \mathcal{J}\right)$.

The power generation by producer $i$ using unit $j$ at node $n$ results in environmental pollution $x^i_{nj}\left(q^i_{nj}\right)$. We assume that $x^i_{nj}(0) = 0$, $x^i_{nj}\left(q^i_{nj}\right)$ is differentiable and is increasing in $q^i_{nj}$. Environmental pollution results in a *negative externality* of $E_n\left(x_n\left(\left[q^i_{nj}\right]^i_j\right)\right)$, which is assumed to be differentiable, increasing and convex in $x_n$.

We model consumers as a continuum due to their large numbers compared to producers, and denote the aggregate power consumption of consumers at node $n$ by

$$d_n \geq 0, \tag{2}$$

and we denote by $\tilde{U}_n(d_n)$ the utility of consuming $d_n$, which we assume is twice-differentiable and non-decreasing in $d_n$. In addition, we make the reasonable assumption that the marginal utility satisfies

$$\frac{\partial}{\partial d_n} \frac{\partial \tilde{U}_n(d_n)}{\partial d_n} \leq 0, \tag{3}$$

i.e., $\tilde{U}_n(d_n)$ is concave in $d_n$. To maintain power balance in the system, the total power generation must equal the total power consumption,

$$q = d. \tag{4}$$

Transmission lines carry electric power from nodes with excess generation to nodes with shortage of generation. The power flow from node $n$ through transmission line $l$ can be expressed as $H_{ln}\left(q_n - d_n\right)$, where $H_{ln}$ is the power transfer distribution factor, which is assumed to be known. The power flow for transmission line $l$ must not exceed its capacity $f_l \geq 0$, i.e.,

$$-f_l \leq \sum_n H_{ln}\left(q_n - d_n\right) \leq f_l. \tag{5}$$

In the absence of transmission line constraints, we would have a single price region $|\mathcal{N}| = 1$.

### 3.1. Socially optimal equilibrium (our desired solution)

Based on the above model, we now consider maximisation of the social welfare in the electricity spot market. Maximisation of the social welfare entails maximising the total utility minus total costs including externalities as

$$\text{maximise}_{\left(\left[q^i_{nj}\right]^i_{nj}, \left[d_n\right]_n\right)} \sum_{n'} \left(\tilde{U}_{n'}(d_{n'}) - \sum_{i'j'} C^{i'}_{n'j'}\left(q^{i'}_{n'j'}\right) - E_{n'}(x_{n'})\right) \tag{6}$$

subject to the constraints (1), (2), (4), and (5). As it is only the utility function $\tilde{U}_n(d_n)$ that depends on $d_n$, and is maximised, we can express the optimal utility as

$$U\left(\left[q_n\right]_n\right) = \max_{\left[d_n\right]_n} \sum_{n'} \tilde{U}_{n'}(d_{n'}) \tag{7}$$

subject to all the constraints except the generation capacity constraints (2), (4) and (5). Observe that $U := U\left(\left[q_n\right]_n\right)$ because the constraints depend only on $\left[q^i_{nj}\right]_n$ and $\left[d_n\right]_n$ not on individual $\left[q_n\right]^i_{nj}$, and $\left[d_n\right]_n$ is the maximiser. As the power balance constraint (4) is always binding, $\sum_{n'} \partial d_{n'}/\partial q_n = 1$. Recall that $\tilde{U}_n(d_n)$ is non-decreasing in $d_n$ for every $n$, and hence $U\left(\left[q_n\right]_n\right)$ is non-decreasing in $q_n$. Also, $\partial U\left(\left[q_{n''}\right]_n\right)/\partial q_n$ is continuous and piecewise differentiable in $q_{n'}$ for every $n$ and $n'$, and thus as the marginal utility is non-decreasing from (3) we have

$$\frac{\partial}{\partial q_{n'}} \frac{\partial U\left(\left[q_{n''}\right]_n\right)}{\partial q_n} \leq 0. \tag{8}$$

Here, in order to allow the derivative of $\partial U\left(\left[q_{n''}\right]_n\right)/\partial q_n$ with respect to $q_{n'}$ to exist everywhere, we have only considered its right-hand derivative[4] at every $\left[q_{n''}\right]_n$ without explicitly denoting it. We will follow this as a convention for all piecewise differentiable functions.

Including the expression of optimal utility from (7) in the social welfare maximisation in (6), we can express the socially *optimal* generation levels as

$$\left[q^{i*}_{nj}\right]^i_{nj} \in \arg\max_{\left[q^i_{nj}\right]^i_{nj}} \left( U\left(\left[q_{n'}\right]_{n'}\right) - \sum_{n'} \left( \sum_{i'j'} C^{i'}_{n'j'}\left(q^{i'}_{n'j'}\right) + E_{n'}(x_{n'}) \right) \right) \tag{9}$$

subject to the generation capacity constraints (1). Recall that constraints (2), (4), and (5) are implicit in $U\left(\left[q_n\right]_n\right)$. Accordingly, for node $n$, producer $i$ and unit $j$, we can characterise the optimal generation levels $q^{i*}_{nj}$ by

$$q^{i*}_{nj} \in \begin{cases} \{0\} & \text{if } \left( \frac{\partial U([q_{n'}]_{n'})}{\partial q_n} - \frac{\partial C^i_{nj}\left(q^i_{nj}\right)}{\partial q^i_{nj}} - \frac{\partial E_n(x_n)}{\partial x_n}\frac{\partial x^i_{nj}}{\partial q^i_{nj}} \right)\Bigg|_{\substack{q^i_{nj}=0,\left[q^i_{nj'}\right]_{j'\neq j}=\left[q^{i*}_{nj'}\right]_{j'\neq j},\\ \left[q^{i'}_{nj'}\right]^{i'\neq i}_{j'}=\left[q^{i'*}_{nj'}\right]^{i'\neq i}_{j'},\left[q^{i'}_{n'j'}\right]^{i'}_{n'\neq nj'}=\left[q^{i'*}_{n'j'}\right]^{i'}_{n'\neq nj'}}} < 0, \\ \left\{k^i_{nj}\right\} & \text{if } \left( \frac{\partial U([q_{n'}]_{n'})}{\partial q_n} - \frac{\partial C^i_{nj}\left(q^i_{nj}\right)}{\partial q^i_{nj}} - \frac{\partial E_n(x_n)}{\partial x_n}\frac{\partial x^i_{nj}}{\partial q^i_{nj}} \right)\Bigg|_{\substack{q^i_{nj}=k^i_{nj},\left[q^i_{nj'}\right]_{j'\neq j}=\left[q^{i*}_{nj'}\right]_{j'\neq j},\\ \left[q^{i'}_{nj'}\right]^{i'\neq i}_{j'}=\left[q^{i'*}_{nj'}\right]^{i'\neq i}_{j'},\left[q^{i'}_{n'j'}\right]^{i'}_{n'\neq nj'}=\left[q^{i'*}_{n'j'}\right]^{i'}_{n'\neq nj'}}} > 0, \\ \left[0, k^i_{nj}\right] & \text{s.t. } \left( \frac{\partial U([q_{n'}]_{n'})}{\partial q_n} - \frac{\partial C^i_{nj}\left(q^i_{nj}\right)}{\partial q^i_{nj}} - \frac{\partial E_n(x_n)}{\partial x_n}\frac{\partial x^i_{nj}}{\partial q^i_{nj}} \right)\Bigg|_{\left[q^{i'}_{n'j'}\right]^{i'}_{n'j'}=\left[q^{i'*}_{n'j'}\right]^{i'}_{n'j'}} = 0 \text{ otherwise.} \end{cases} \tag{10}$$

### 3.2. Competitive equilibrium (ignoring pollution damage)

In practice, the electricity spot market is operated by an Independent System Operator (ISO), whose objective, as set by the energy regulator, is to maximise the social welfare in the electricity spot market. Nonetheless, in general, the ISO suffers from *information asymmetry*, i.e., it has incomplete information about the constraints and the preferences of the producers and the consumers. First, the ISO may not observe $\left[q^i_{nj}\right]^i_{nj}$ or $\left[x^i_{nj}\right]^i_{nj}$, but can only observe the total output $\left[q^i_n\right]^i_n$ and $\left[x^i_n\right]^i_n$. Second, to maximise social welfare, the ISO has to rely on every producer $i$ to declare their total cost functions $\tilde{C}^i_n\left(q^i_n\right)$ for every node $n$. We consider that $\tilde{C}^i_n\left(q^i_n\right)$ is continuous, piecewise differentiable, increasing and convex in $q^i_n$, and $\tilde{C}^i_n(0) = 0$. It is up to producer $i$ to construct $\tilde{C}^i_n\left(q^i_n\right)$ given $C^i_{nj}\left(q^i_{nj}\right)$. Third, although the ISO may know the total available generation capacity $k^i_n$ of producer $i$ at node $n$, it may not know the available generation capacity for each unit, which would only allow it to formulate a constraint on feasible generation levels per producer and node (as opposed to the generation capacity constraints for individual technologies (1)),

$$0 \leq q^i_n \leq k^i_n. \tag{11}$$

Fourth, the ISO has to rely on consumers to declare their utility functions $\tilde{U}_n(d_n)$ for every node $n$. Finally, observe that the pollution $x^i_n$ and the externality $E_n(x_n)$ are unit dependent, and thus they cannot be expressed as a function of $q^i_n$.

Due to this information obscurity, instead of computing the socially optimal generation levels from (9) the ISO approximates the social welfare maximisation and computes the *spot market* generation levels as

$$\left[q^{i\dagger}_n\right]^i_n \in \arg\max_{\left[q^i_n\right]^i_n} \left( U\left(\left[q_{n'}\right]_{n'}\right) - \sum_{n'i'} \tilde{C}^{i'}_{n'}\left(q^{i'}_{n'}\right) \right) \tag{12}$$

subject to (11). As there is no other dependence on $d_n$ in the above, we have used $U\left(\left[q_{n'}\right]_{n'}\right)$ from (7). The above allows us to characterise the spot market generation levels as

$$q^{i\dagger}_n \in \begin{cases} \{0\} & \text{if } \left( \frac{\partial U([q_{n'}]_{n'})}{\partial q_n} - \frac{\partial \tilde{C}^i_n(q^i_n)}{\partial q^i_n} \right)\Bigg|_{q^i_n=0,\left[q^{i'}_n\right]^{i'\neq i}=\left[q^{i'\dagger}_n\right]^{i'\neq i},\left[q^{i'}_{n'}\right]^{i'}_{n'\neq n}=\left[q^{i'\dagger}_{n'}\right]^{i'}_{n'\neq n}} < 0, \\ \left\{k^i_n\right\} & \text{if } \left( \frac{\partial U([q_{n'}]_{n'})}{\partial q_n} - \frac{\partial \tilde{C}^i_n(q^i_n)}{\partial q^i_n} \right)\Bigg|_{q^i_n=k^i_n,\left[q^{i'}_n\right]^{i'\neq i}=\left[q^{i'\dagger}_n\right]^{i'\neq i},\left[q^{i'}_{n'}\right]^{i'}_{n'\neq n}=\left[q^{i'\dagger}_{n'}\right]^{i'}_{n'\neq n}} > 0, \\ \left[0, k^i_n\right] & \text{s.t. } \left( \frac{\partial U([q_{n'}]_{n'})}{\partial q_n} - \frac{\partial \tilde{C}^i_n(q^i_n)}{\partial q^i_n} \right)\Bigg|_{\left[q^{i'}_{n'}\right]^{i'}_{n'}=\left[q^{i'\dagger}_{n'}\right]^{i'}_{n'}} = 0 \text{ otherwise.} \end{cases} \tag{13}$$

Based on the above, the ISO can set the price $P_n$ at node $n$ that maximises the social welfare (not accounting for the externality), as

$$P_n\left(\left[q_{n'}\right]_{n'}\right) = \frac{\partial U\left(\left[q_{n'}\right]_{n'}\right)}{\partial q_n} \tag{14}$$

and following from (13) we have

$$q^{i\dagger}_n \in \left[0, k^i_n\right] \text{ if } \frac{\partial \tilde{C}^i_n\left(q^i_n\right)}{\partial q^i_n}\Bigg|_{\left[q^{i'}_{n'}\right]^{i'}_{n'}=\left[q^{i'\dagger}_{n'}\right]^{i'}_{n'}} = P_n\left(\left[q^{\dagger}_{n'}\right]_{n'}\right). \tag{15}$$

As $U\left(\left[q_{n'}\right]_{n'}\right)$ is non-decreasing in $q_n$ for every node $n$, $P_n(\left[q_{n'}\right]_{n'})$ is non-negative. Furthermore, $P_{n'}(\left[q_{n''}\right]_{n''})$ is continuous, piecewise differentiable and non-increasing in $q_n$ due to (8).

For the competitive case, let us consider that every producer $i$ declares their total capacity and total cost function truthfully. To characterise the *competitive* solution, let us first express the minimal total cost function of producer $i$ at node $n$ as

$$C^i_n\left(q^i_n\right) = \min_{\left[\overline{q}^i_{nj}\right]_j} \sum_{j'} C^i_{nj'}\left(\overline{q}^i_{nj'}\right) \tag{16}$$

$$\text{subject to } \overline{q}^i_n = q^i_n \tag{17}$$

[4] The right-hand derivative of function $g(z)$ with respect to $z$ is defined as $\left.\frac{\partial g(z)}{\partial z}\right|_+ = \lim_{h\to 0^+} \frac{g(z+h)-g(z)}{h}$.

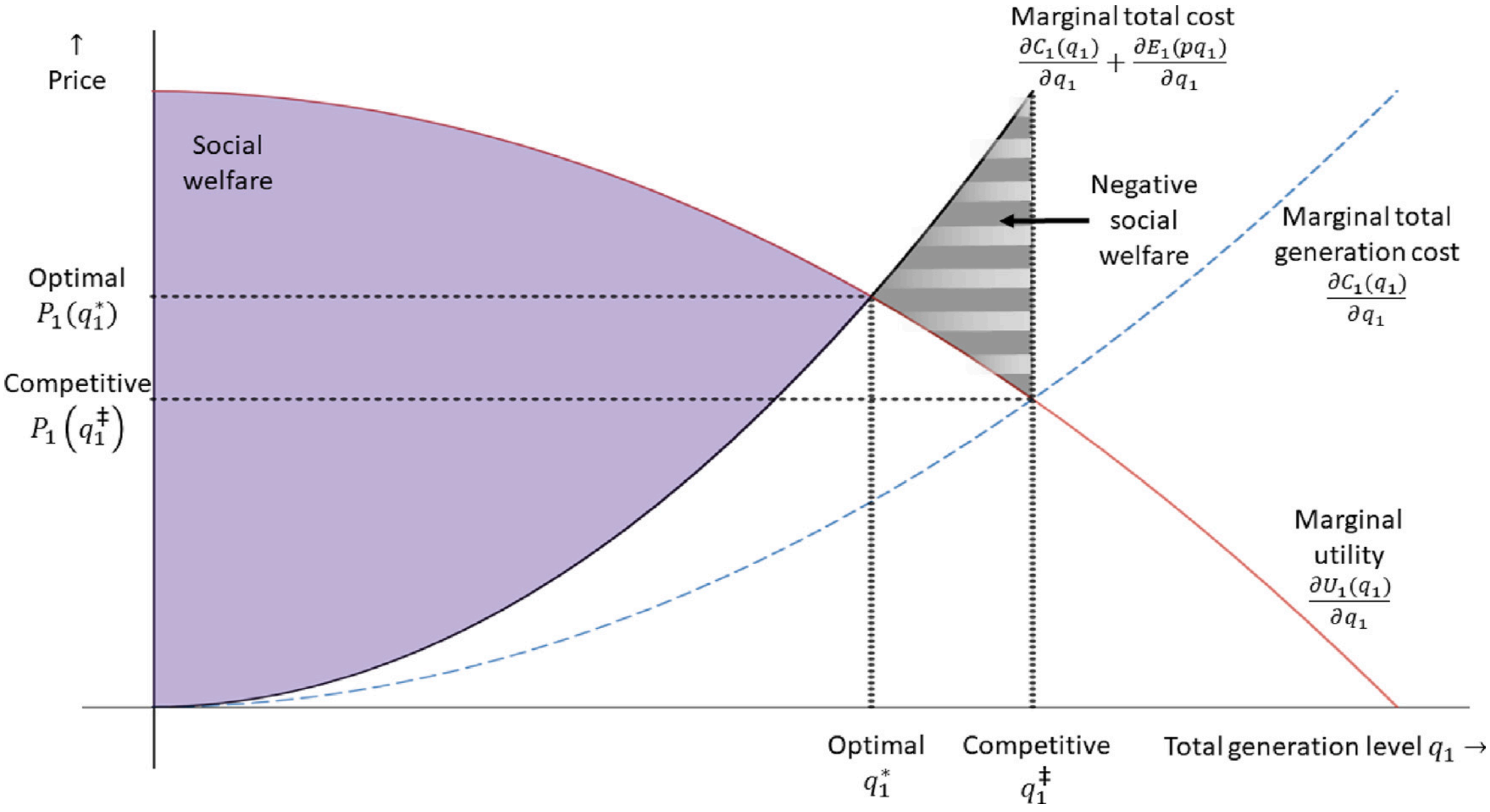

**Fig. 1.** Negative social welfare due to ignoring pollution.

and the generation capacity constraints (1). This formulation allows us to characterise the cost minimising generation levels,

$$\bar{q}^i_{nj}\left(q^i_n\right) \in \begin{cases} \{0\} & \text{if } \left.\dfrac{\partial C^i_{nj}\left(\bar{q}^i_{nj}\right)}{\partial \bar{q}^i_{nj}}\right|_{\bar{q}^i_{nj}=0} > \dfrac{\partial C^i_n\left(\bar{q}^i_n\right)}{\partial \bar{q}^i_n}, \\ \left\{k^i_{nj}\right\} & \text{if } \left.\dfrac{\partial C^i_{nj}\left(\bar{q}^i_{nj}\right)}{\partial \bar{q}^i_{nj}}\right|_{\bar{q}^i_{nj}=k^i_{nj}} < \dfrac{\partial C^i_n(q^i_n)}{\partial q^i_n}, \\ \left[0, k^i_{nj}\right] & \text{s.t. } \dfrac{\partial C^i_{nj}\left(\bar{q}^i_{nj}\right)}{\partial \bar{q}^i_{nj}} = \dfrac{\partial C^i_n(q^i_n)}{\partial q^i_n} \text{ otherwise.} \end{cases} \tag{18}$$

Recall that, for every $n$, $i$ and $j$, $C^i_{nj}\left(q^i_{nj}\right)$ is differentiable, increasing and convex in $q^i_{nj}$. Consequently, $C^i_n\left(q^i_n\right)$ is continuous, piecewise differentiable, increasing and convex in $q^i_n$, and $C^i_n(0) = 0$. Accordingly, under a truthful declaration $\tilde{C}^i_n\left(q^i_1\right) = C^i_n\left(q^i_1\right) = \sum_{j'} C^i_{nj'}\left(q^i_{nj'}\right)$. The competitive generation levels can be characterised as

$$q^{i\ddagger}_n \in \begin{cases} \{0\} & \text{if } \left.\left(\dfrac{\partial U([q_{n'}]_{n'})}{\partial q_n} - \dfrac{\partial C^i_n(q^i_n)}{\partial q^i_n}\right)\right|_{q^i_n=0,\left[q^{i'}_n\right]^{i'\neq i}=\left[q^{i'\ddagger}_n\right]^{i'\neq i},\left[q^{i'}_{n'}\right]^{i'}_{n'\neq n}=\left[q^{i'\ddagger}_{n'}\right]^{i'}_{n'\neq n}} < 0, \\ \left\{k^i_n\right\} & \text{if } \left.\left(\dfrac{\partial U([q_{n'}]_{n'})}{\partial q_n} - \dfrac{\partial C^i_n(q^i_n)}{\partial q^i_n}\right)\right|_{q^i_n=k^i_n,\left[q^{i'}_n\right]^{i'\neq i}=\left[q^{i'\ddagger}_n\right]^{i'\neq i},\left[q^{i'}_{n'}\right]^{i'}_{n'\neq n}=\left[q^{i'\ddagger}_{n'}\right]^{i'}_{n'\neq n}} > 0, \\ \left[0, k^i_n\right] & \text{s.t. } \left.\left(\dfrac{\partial U([q_{n'}]_{n'})}{\partial q_n} - \dfrac{\partial C^i_n(q^i_n)}{\partial q^i_n}\right)\right|_{\left[q^{i'}_{n'}\right]^{i'}_{n'}=\left[q^{i'\ddagger}_{n'}\right]^{i'}_{n'}} = 0 \text{ otherwise.} \end{cases} \tag{19}$$

As an illustration, consider a power system with a single node, $\mathcal{N} = \{1\}$ and consequently, no transmission lines, $\mathcal{L} = \emptyset$. Accordingly, based on the power balance constraint (4) and the definition of optimal utility in (7), we have $U(q_1) = \tilde{U}_1(d_1)$. For simplicity, consider a linear pollution function $x^i_{1j} = pq^i_{1j}$ $\forall i \in \mathcal{I}, j \in \mathcal{J}$, where $p > 0$ is a constant independent of $i$ and $j$. The resulting externality is $E_1(x_1) = E_1(pq_1)$, and depends only on the total generation level $q_1$. The optimal generation levels in this case would be

$$\left[q^{i*}_{1j}\right]^i_j \in \arg\max_{\left[q^i_{1j}\right]^i_j} \left( U(q_1) - \sum_{i'j'} C^{i'}_{1j'}\left(q^{i'}_{1j'}\right) - E_1(pq_1) \right) \tag{20}$$

subject to the generation capacity constraints (1). Observe that, of all of the terms in the objective function, it is only the generation cost function that is parametrised by $i$ and $j$. In order to eliminate this dependence and express the above only in terms of $q_1$, we can define the minimal total cost function of generation level $q_1$ as

$$C_1(q_1) = \min_{\left[\bar{q}^i_{1j}\right]^i_j} \sum_{i'j'} C^{i'}_{1j'}\left(\bar{q}^{i'}_{1j'}\right) \tag{21}$$

$$\text{subject to } \bar{q}_1 = q_1 \tag{22}$$

and (1). The cost minimising generation levels at $q_1$ can then be characterised by

$$\bar{q}^i_{1j}(q_1) \in \begin{cases} \{0\} & \text{if } \left.\dfrac{\partial C^i_{1j}\left(\bar{q}^i_{1j}\right)}{\partial \bar{q}^i_{1j}}\right|_{\bar{q}^i_{1j}=0} > \dfrac{\partial C_1(q_1)}{\partial q_1}, \\ \left\{k^i_{1j}\right\} & \text{if } \left.\dfrac{\partial C^i_{1j}\left(\bar{q}^i_{1j}\right)}{\partial \bar{q}^i_{1j}}\right|_{\bar{q}^i_{1j}=k^i_{1j}} < \dfrac{\partial C_1(q_1)}{\partial q_1}, \\ \left[0, k^i_{1j}\right] & \text{s.t. } \dfrac{\partial C_1(q_1)}{\partial q_1} = \dfrac{\partial C^i_{1j}\left(\bar{q}^i_{1j}\right)}{\partial \bar{q}^i_{1j}} \text{ otherwise.} \end{cases} \tag{23}$$

Consequently, $C_1\left(q_1\right)$ is continuous, piecewise differentiable, increasing and convex in $q_1$. Also, assume that every producer declares their total capacity and total cost function truthfully, i.e., $k^i_1 = \sum_{j'} k^i_{1j'}$ $\forall i \in \mathcal{I}$ and $\tilde{C}^i_1\left(q^i_1\right) = C^i_1\left(q^i_1\right) = \sum_{j'} C^i_{1j'}\left(q^i_{1j'}\right)$ $\forall i \in \mathcal{I}$. In Fig. 1, we illustrate the marginal total generation cost $\partial C_1(q_1)/\partial q_1$, the marginal total cost $\partial C_1(q_1)/\partial q_1 + \partial E_1(pq_1)/\partial q_1$ and the marginal consumer utility $\partial U_1(q_1)/\partial q_1$. From the intersection of the marginal total cost and the marginal consumer utility, we derive the optimal $q^*_1$. Given $q^*_1$, the optimal generation levels for producer $i$ and unit $j$ can be determined from (23) as $q^{i*}_{1j} = \bar{q}^i_{1j}(q^*_1)$. We also illustrate the corresponding price $P_1\left(q^*_1\right)$ from (14). Observe that we can determine an *optimal price* for this example only because the externality $E_1(pq_1)$ can be represented as a function of the total generation $q_1$. The solid shaded region illustrates the resulting social welfare in the optimal case. This is the well-known optimal theory leading to marginal cost pricing from the producer supply function to match the marginal willingness to pay from the demand-side.

We also illustrate the competitive $q^{\ddagger}_1$ which we derive from the intersection of the marginal total generation cost and the marginal consumer utility, and the corresponding price $P_1\left(q^{\ddagger}_1\right)$. As the externality $E_1(pq_1)$ is not accounted for, $q^{\ddagger}_1 > q^*_1$. Consequently, the competitive price is lower than the optimal price $P_1\left(q^{\ddagger}_1\right) < P_1\left(q^*_1\right)$. In addition to the optimal social welfare, ranging from $q^*_1$ to $q^{\ddagger}_1$, there is a striped region of *negative social welfare* that will be realised in the competitive case where the marginal total cost exceeds the marginal consumer utility. This illustrates that the competitive solution does not maximise social welfare.

### 3.3. Equilibrium under imperfect competition (ignoring pollution damage)

We have so far considered the optimal generation levels $\left[q_n^{i*}\right]_n^i$ and the spot market generation levels $\left[q_n^{i\dagger}\right]_n^i$. With imperfect competition due to an oligopoly, the generation levels may differ from $\left[q_n^{i*}\right]_n^i$ and $\left[q_n^{i\dagger}\right]_n^i$.

In practice, producer $i$ has a prediction of the cost functions $\tilde{C}_n^{i'}\left(q_n^{i'}\right)$ and the generation capacities $\left[k_n^{i'}\right]_n^{i'\neq i}$ for every other producer $i' \neq i$ at each node $n$, and of the utility function $U\left(\left[q_{n'}\right]_{n'}\right)$. In the Bayesian Nash equilibrium, the predictions would be correct, and to maximise their profit, producer $i$ can use (14) to manipulate $P_n\left(\left[q_{n'}\right]_{n'}\right)$. Accordingly, their desired *oligopolistic* generation levels are

$$\left[q_{nj}^{i\#}\right]_{nj} \in \arg\max_{\left[q_{nj}^i\right]_{nj}} \sum_{n'} \left( P_{n'}\left(\left[q_{n''}\right]_{n''}\right) q_{n'}^i - \sum_{j'} C_{n'j'}^i\left(q_{n'j'}^i\right) \right) \tag{24}$$

subject to the generation capacity constraints (1). Observe that this maximisation does not include the externality $E_n(x_n)$, as by definition, individual producers and consumers do not account for the externalities in their objective functions.

Including the minimal total cost function (16), the oligopolistic generation levels of producer $i$ can be expressed as

$$\left[q_n^{i\#}\right]_n \in \arg\max_{\left[q_n^i\right]_n} \sum_{n'} \left( P_{n'}\left(\left[q_{n''}\right]_{n''}\right) q_{n'}^i - C_{n'}^i\left(q_{n'}^i\right) \right) \tag{25}$$

subject to the generation capacity constraints per producer per node (11). This allows us to characterise the oligopolistic generation levels,

$$q_n^{i\#} \in \begin{cases} \{0\} & \text{if } \left( \displaystyle\sum_{n'} \frac{\partial P_{n'}\left(\left[q_{n''}\right]_{n''}\right)}{\partial q_n} q_{n'}^i + P_n\left(\left[q_{n'}\right]_{n'}\right) - \frac{\partial C_n^i\left(q_n^i\right)}{\partial q_n^i} \right)\Bigg|_{q_n^i=0,\left[q_{n'}^i\right]_{n'\neq n}=\left[q_{n'}^{i\#}\right]_{n'\neq n}} < 0, \\ \left\{k_{nj}^i\right\} & \text{if } \left( \displaystyle\sum_{n'} \frac{\partial P_{n'}\left(\left[q_{n''}\right]_{n''}\right)}{\partial q_n} q_{n'}^i + P_n\left(\left[q_{n'}\right]_{n'}\right) - \frac{\partial C_n^i\left(q_n^i\right)}{\partial q_n^i} \right)\Bigg|_{q_n^i=k_n^i,\left[q_{n'}^i\right]_{n'\neq n}=\left[q_{n'}^{i\#}\right]_{n'\neq n}} > 0, \\ \left[0, k_{nj}^i\right] & \text{s.t. } \left( \displaystyle\sum_{n'} \frac{\partial P_{n'}\left(\left[q_{n''}\right]_{n''}\right)}{\partial q_n} q_{n'}^i + P_n\left(\left[q_{n'}\right]_{n'}\right) - \frac{\partial C_n^i\left(q_n^i\right)}{\partial q_n^i} \right)\Bigg|_{\left[q_{n'}^i\right]_{n'}=\left[q_{n'}^{i\#}\right]_{n'}} = 0 \text{ otherwise.} \end{cases} \tag{26}$$

Comparing the above to the optimal generation levels $q_{nj}^{i*}$ from (10) we can observe that the oligopolistic generation levels differ from the optimal generation levels, i.e., $q_n^{i\#} \neq q_n^{i*}$.

Producer $i$ has an incentive to realise their oligopolistic generation level $q_n^{i\#}$ in the spot market, i.e., $q_n^{i\dagger} = q_n^{i\#}$ where $q_n^{i\dagger}$ is defined in (19). To achieve this, from the definition of price in (14), we can deduce that it would be sufficient for producer $i$ to declare a cost function $\tilde{C}_n^i\left(q_n^i\right)$ that obeys

$$\frac{\partial \tilde{C}_n^i\left(q_n^i\right)}{\partial q_n^i} = \frac{\partial C_n^i\left(q_n^i\right)}{\partial q_n^i} - \sum_{n'} \frac{\partial P_{n'}\left(\left[q_{n''}\right]_{n''}\right)}{\partial q_n} q_{n'}^i \geq \frac{\partial C_n^i\left(q_n^i\right)}{\partial q_n^i}. \tag{27}$$

Observe that $\tilde{C}_n^i\left(q_n^i\right)$ is by construction continuous, piecewise differentiable, increasing and convex in $q_n^i$, as required above. We can obtain an analytic expression for $\tilde{C}_n^i\left(q_n^i\right)$ by recalling that $C_{nj}^i(0) = 0\ \forall n \in \mathcal{N}, i \in \mathcal{I}, j \in \mathcal{J}$, enforcing $\tilde{C}_n^i(0) = 0\ \forall n \in \mathcal{N}, i \in \mathcal{I}$, and by integration with respect to $q_n^i$,

$$\begin{aligned} \tilde{C}_n^i\left(q_n^i\right) = {} & C_n^i\left(q_n^i\right) + U\left(\left[q_{n'}\right]_{n'}\right) - U\left(\left[q_{n'}\right]_{n'<n}, q_n - q_n^i, \left[q_{n'}\right]_{n'>n}\right) \\ & - \sum_{n'} \left( P_{n'}\left(\left[q_{n''}\right]_{n''}\right) - P_{n'}\left(\left[q_{n''}\right]_{n''<n}, q_n - q_n^i, \left[q_{n''}\right]_{n''>n}\right)\right) q_{n'}^i \\ & - P_n\left(\left[q_{n'}\right]_{n'<n}, q_n - q_n^i, \left[q_{n'}\right]_{n'>n}\right) q_n^i \geq C_n^i\left(q_n^i\right). \end{aligned} \tag{28}$$

From the definition of price in (15), we can see that declaring a marginal cost $\partial \tilde{C}_n^i\left(q_n^i\right)/\partial q_n^i > \partial C_n^i\left(q_n^i\right)/\partial q_n^i$ for every $q_n^i$ may decrease $q_n^{i\dagger}$, but may increase $P_{n'}\left(\left[q_{n''}\right]_{n''}\right)$, and hence the overall profit. The ability of producers to control prices to increase profits, i.e., their *market power* can be quantified as $-\sum_{n'} \frac{\partial P_{n'}\left(\left[q_{n''}\right]_{n''}\right)}{\partial q_n} q_{n'}^i$ for producer $i$ and node $n$.

To illustrate market power, consider again the example described previously, and the total cost function $\tilde{C}_1^i\left(q_n^i\right)$ declared by producer $i$ according to (28). Observe that, in the spot market generation levels in (12), of all the terms in the objective function, it is only the declared generation cost function that is parametrised by $i$. In order to eliminate this dependence and express (12) only in terms of $q_1$, we can define the minimal total declared cost function as

$$\tilde{C}_1(q_1) = \min_{\left[\overline{q}_1^i\right]^i} \sum_{i'} \tilde{C}_1^{i'}\left(\overline{q}_1^{i'}\right) \tag{29}$$

$$\text{subject to } \overline{q}_1 = q_1 \tag{30}$$

and the generation capacity constraints (11). This allows us to characterise the cost minimising generation levels,

$$\overline{q}_1^i\left(q_1\right) \in \begin{cases} \{0\} & \text{if } \left.\dfrac{\partial \tilde{C}_1^i\left(\overline{q}_1^i\right)}{\partial \overline{q}_1^i}\right|_{\overline{q}_1^i=0} > \dfrac{\partial \tilde{C}_1(q_1)}{\partial q_1}, \\ \left\{k_1^i\right\} & \text{if } \left.\dfrac{\partial \tilde{C}_1^i\left(\overline{q}_1^i\right)}{\partial \overline{q}_1^i}\right|_{\overline{q}_1^i=k_1^i} < \dfrac{\partial \tilde{C}_1(q_1)}{\partial q_1}, \\ \left[0, k_1^i\right] & \text{s.t. } \dfrac{\partial \tilde{C}_1^i\left(\overline{q}_1^i\right)}{\partial \overline{q}_1^i} = \dfrac{\partial \tilde{C}_1(q_1)}{\partial q_1} \text{ otherwise.} \end{cases} \tag{31}$$

Consequently, $\tilde{C}_1\left(q_1\right)$ is continuous, piecewise differentiable, increasing and convex in $q_1$. In Fig. 2, we illustrate the marginal total declared cost $\partial \tilde{C}_1(q_1)/\partial q_1$. We derive the optimal $q_1^*$, the competitive $q_1^{\ddagger}$ and the oligopolistic $q_1^{\#}$, and the corresponding prices from (14). As the marginal total declared cost is greater than the marginal total generation cost $\partial \tilde{C}_1(q_1)/\partial q_1 > \partial C_1(q_1)/\partial q_1$, $q_1^{\#} < q_1^{\ddagger}$, and the oligopolistic price is higher than the competitive price $P_1\left(q_1^{\#}\right) > P_1\left(q_1^{\ddagger}\right)$. Also, $q_1^{\#}$ can be equal to, greater than or less than $q_1^*$, depending on the relative values of marginal costs and utility. If $q_1^{\#} = q_1^*$ then social welfare is maximised. Alternatively, $q_1^{\#} > q_1^*$ results in a region of negative social welfare as seen in Fig. 1. In Fig. 2, we illustrate the case where $q_1^{\#} < q_1^*$. Ranging from $q_1^{\#}$ to $q_1^*$, there is a striped region of *social welfare loss*, i.e., social welfare that would not be realised due to the declared cost function. The total profit of all producers is illustrated by the part of the solid shaded region below the price. Comparing Fig. 1, which shows the competitive solution, and Fig. 2, which shows the oligopolistic solution, we can see that the total profit of all producers increases even though social welfare decreases. As both the optimal ($q_1^*$) and oligopolistic ($q_1^{\#}$) generation levels are less than the competitive generation levels ($q_1^{\ddagger}$), one could conclude that market power inadvertently alleviates pollution. However, the extent of market power cannot be controlled by the ISO and therefore the resulting solution does not maximise social welfare. Furthermore, regulating market power with a price cap is not optimal, as we show in the next section.

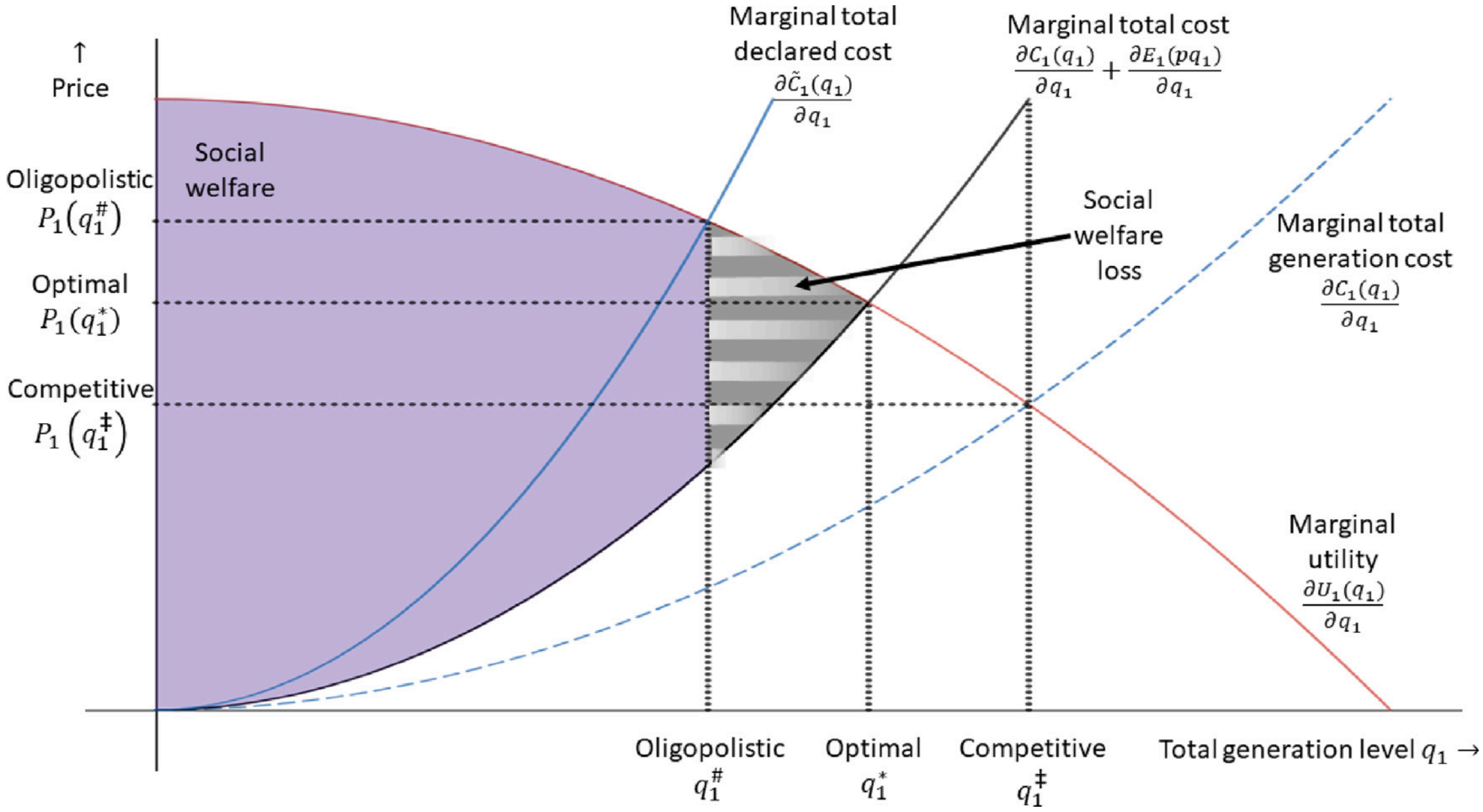

**Fig. 2.** Social welfare loss due to market power.

**Table 1**
Summary of conditions for optimal, competitive and oligopolistic generation levels.

| | $\frac{\partial U([q_{n'}]_{n'})}{\partial q_n} = P_n([q_{n'}]_{n'})$ | $-\frac{\partial C^i_{nj}(q^i_{nj})}{\partial q^i_{nj}}$ | $-\frac{\partial E_n(x_n)}{\partial x_n}\frac{\partial x^i_{nj}}{\partial q^i_{nj}}$ | $+\sum_{n'}\frac{\partial P_{n'}([q_{n''}]_{n''})}{\partial q_n} q^i_{n'}$ |
|---|---|---|---|---|
| Optimal | ✓ | ✓ | ✓ | |
| Competitive | ✓ | ✓ | | |
| Oligopolistic | ✓ | ✓ | | ✓ |

To summarise, observe that the optimal, competitive and oligopolistic generation levels can be determined from an expression of the form

$$q^{i\square}_{nj} \in \begin{cases} \{0\} & \text{if } f^i_{nj}\left(\left[q^{i'}_{n'j'}\right]^{i'}_{n'j'}\right)\Big|_{q^i_{nj}=0,\left[q^i_{nj'}\right]_{j'\neq j}=\left[q^{i\square}_{nj'}\right]_{j'\neq j},\ \left[q^{i'}_{nj'}\right]^{i'\neq i}_{j'}=\left[q^{i'\square}_{nj'}\right]^{i'\neq i}_{j'},\ \left[q^{i'}_{n'j'}\right]^{i'}_{n'\neq nj'}=\left[q^{i'\square}_{n'j'}\right]^{i'}_{n'\neq nj'}} < 0, \\ \left\{k^i_{nj}\right\} & \text{if } f^i_{nj}\left(\left[q^{i'}_{n'j'}\right]^{i'}_{n'j'}\right)\Big|_{q^i_{nj}=k^i_{nj},\left[q^i_{nj'}\right]_{j'\neq j}=\left[q^{i\square}_{nj'}\right]_{j'\neq j},\ \left[q^{i'}_{nj'}\right]^{i'\neq i}_{j'}=\left[q^{i'\square}_{nj'}\right]^{i'\neq i}_{j'},\ \left[q^{i'}_{n'j'}\right]^{i'}_{n'\neq nj'}=\left[q^{i'\square}_{n'j'}\right]^{i'}_{n'\neq nj'}} > 0, \\ \left[0, k^i_{nj}\right] & \text{s.t. } f^i_{nj}\left(\left[q^{i'}_{n'j'}\right]^{i'}_{n'j'}\right)\Big|_{\left[q^{i'}_{n'j'}\right]^{i'}_{n'j'}=\left[q^{i'\square}_{n'j'}\right]^{i'}_{n'j'}} = 0 \text{ otherwise} \end{cases} \tag{32}$$

where the superscript $\square$ is a placeholder for $*$, $\ddagger$ and $\#$ respectively and the terms contained in $f^i_{nj}\left(\left[q^{i'}_{n'j'}\right]^{i'}_{n'j'}\right)$ is summarised in Table 1. Here, recall that the marginal utility is equal to the price from (14). Note that the equation above represents individual generation levels $\left[q^i_{nj}\right]^{i\square}_{nj}$ even for the competitive and oligopolistic case. The individual generation levels $\left[q^i_{nj}\right]^{i\square}_{nj}$ can be obtained from the corresponding total levels $\left[q^i_n\right]^{i\square}_n$ from (18) by comparing their marginal costs to the marginal total cost function.

## 4. Mitigating pollution damage and market power

### 4.1. Price caps

As we have shown previously, in Section 3.3, producers' market power causes a deviation from the ISO's desired competitive solution. In practice, ISOs try to alleviate the problem of market power by imposing a price cap $\overline{P}_n$ at every node $n$. Price caps face the drawbacks of not addressing externalities and not completely eliminating market power. In what follows, we will highlight the latter.

Price caps are typically constant over time, sometimes linked to fuel costs. In the presence of a price cap, the price at node $n$ becomes the minimum of the price cap and the price from (14), i.e.,

$$P_n\left(\left[q_{n'}\right]_{n'}\right) = \min\left\{\frac{\partial U\left(\left[q_{n'}\right]_{n'}\right)}{\partial d_n}, \overline{P}_n\right\}. \tag{33}$$

The cost declared by producer $i$ at bus $n$ as a function of total generation $q^i_n$ under the price cap $\overline{P}_n$ is as follows.

1. $\tilde{C}^i_n\left(q^i_n\right)$ from (28) if $0 \leq q^i_n < \hat{q}^i_n$ s.t. $\partial\tilde{C}^i_n/\partial q^i_n\big|_{q^i_n=\hat{q}^i_n} = \overline{P}_n$
   This is because producer $i$ may exercise market power below the price cap.
   In this region, the price cap does not affect market power.
2. $\overline{P}_n$ if $\hat{q}^i_n \leq q^i_n \leq \tilde{q}^i_n$ s.t. $\partial\tilde{C}^i_n/\partial q^i_n\big|_{q^i_n=\hat{q}^i_n} = \overline{P}_n$, $\partial C^i_n/\partial q^i_n\big|_{q^i_n=\tilde{q}^i_n} = \overline{P}_n$
   where $C^i_n$ is from (16)
   This is because producer $i$ will still make a positive profit if the price is higher than their true marginal cost, even if it is less than the profit maximising price. If instead the marginal declared cost is greater than the price cap, producer $i$ will not be able to generate above the price cap.
   It is in this region that the price cap partially mitigates market power.
3. $C^i_n\left(q^i_n\right)$ from (16) if $\tilde{q}^i_n < q^i_n$ s.t. $\partial C^i_n/\partial q^i_n\big|_{q^i_n=\tilde{q}^i_n} = \overline{P}_n$
   This is because producer $i$ would not profit if the price is lower than their true marginal cost otherwise producer $i$ will incur a loss.
   As $C^i_n\left(q^i_n\right)$ is convex in $q^i_n$, the marginal cost for $q^i_n > \tilde{q}^i_n$ will be greater than the price cap, i.e., $\partial C^i_n/\partial q^i_n > \partial C^i_n/\partial q^i_n\big|_{q^i_n=\tilde{q}^i_n} = \overline{P}_n$.
   Therefore, the maximum generation of producer $i$ at bus $n$ is $\tilde{q}^i_n$.

Consider the example developed in Section 3. Fig. 3 represents the case where the price cap is low such that the marginal utility curve intersects the marginal total declared cost curve in region 3 of the curve described above. Therefore, producers would not act to incur losses, from (27) and they will produce a *capped* generation level $q^{i\#}_1$ which is the maximum level such that producers do not incur losses, such that $\partial C_1^i\left(q_1^i\right)/\partial q_1^i\Big|_{q_1^i=q_1^{i\#}} = \overline{P}_1$. Accordingly, the capped generation level is lower than the competitive generation level that the price cap aims to achieve, i.e., $q_1^{i\#} < q_1^{i\ddagger}$. In addition, consumers would wish to consume at a *desired* level $q_1^{\S}$ such that $\partial U(q_1)/\partial q_1\big|_{q_1=q_1^{\S}} = \overline{P}_1$ where $q_1^{\S} \geq q_1^{\ddagger}$. This causes an *involuntary load shedding* by an amount of $q_1^{\S} - q_1^{\#}$ because generation may be less than the desired consumption and thus, price caps have an effect similar to market power. To avoid this undesired effect, price caps are, in practice, set to be very high. In Fig. 3, we illustrate the case above with the price cap $\overline{P}_1$, the marginal total declared cost $\partial\tilde{C}_1/\partial q_1$, the capped oligopolistic $q_1^{\#}$, the desired $q_1^{\S}$ and the involuntary load shedding $q_1^{\S} - q_1^{\#}$.

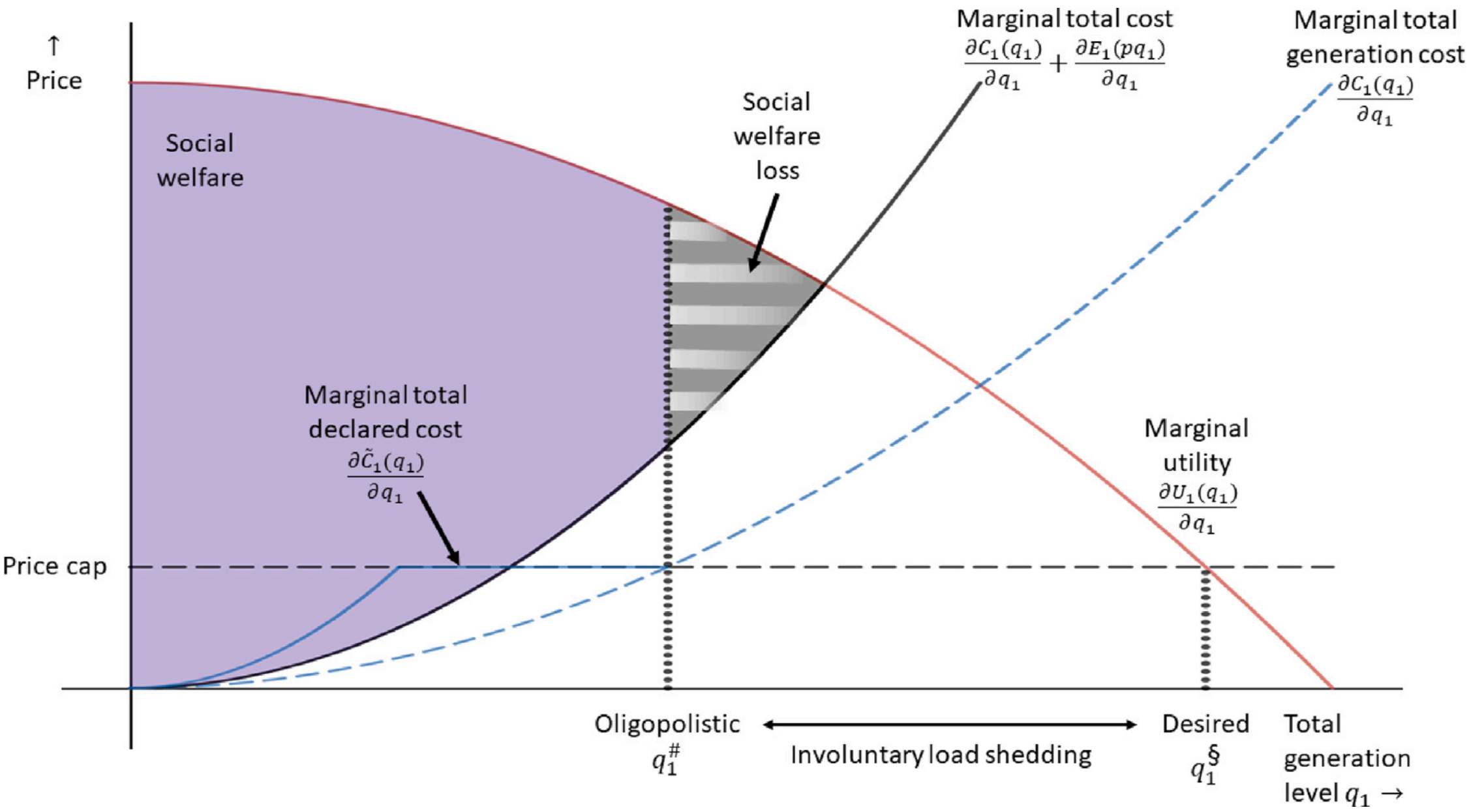

**Fig. 3.** Social welfare loss and involuntary load shedding given a low price cap.

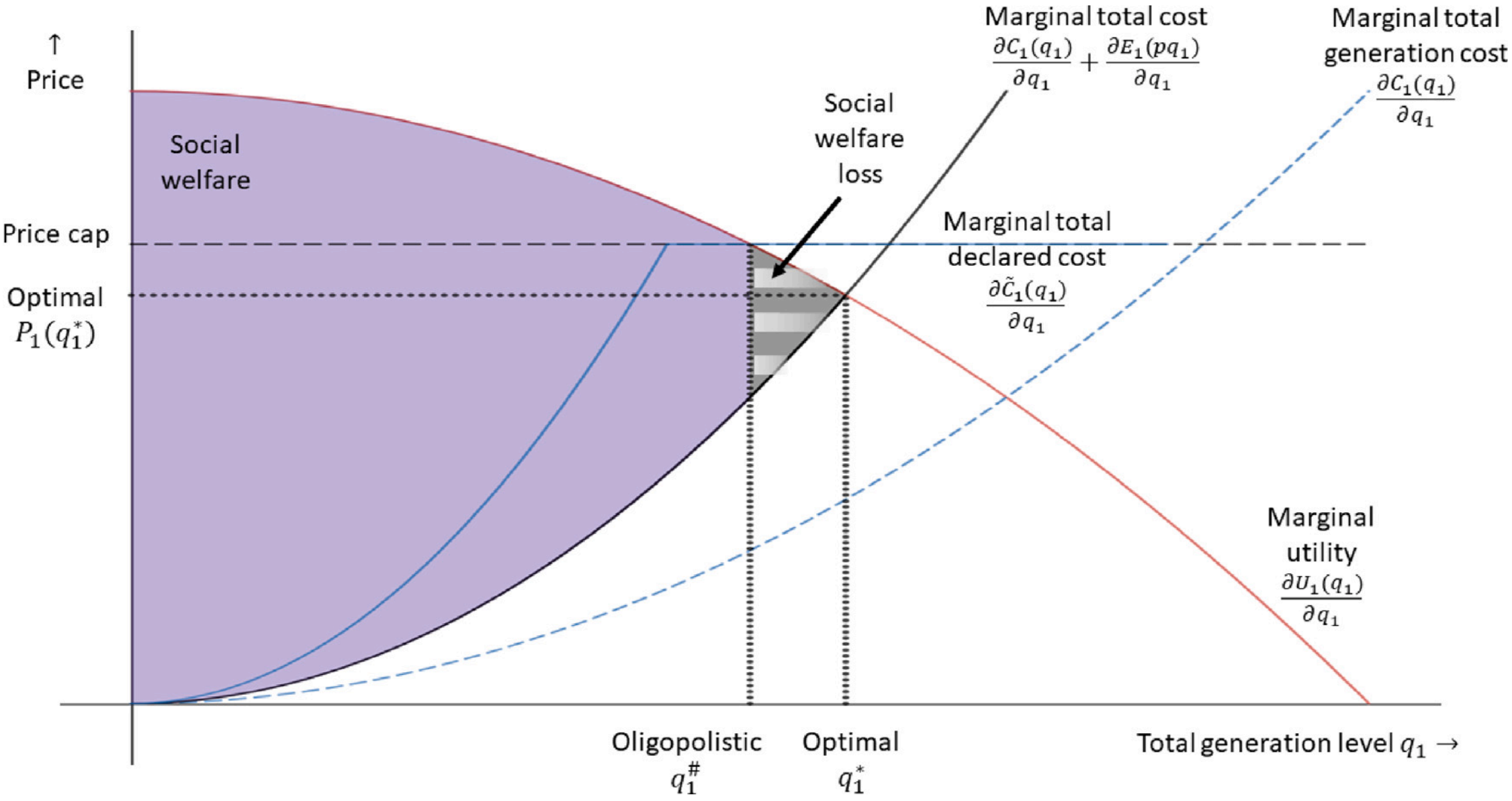

**Fig. 4.** Social welfare loss due to market power given a high price cap.

Fig. 4 considers the more likely case where the price cap is high. In this case, we consider that the marginal utility curve intersects the marginal total declared cost curve in region 2 of the curve described above. Comparing the marginal total declared cost $\partial\tilde{C}_1/\partial q_1$ in Fig. 2 to that in Fig. 4, we can see that producer $i$ can exercise market power by declaring a false cost as if there was no intervention where $\partial\tilde{C}_1^i\left(q_1^i\right)/\partial q_1^i\Big|_{q_1^i=q_1^{i\ddagger}} > \partial C_1^i\left(q_1^i\right)/\partial q_1^i\Big|_{q_1^i=q_1^{i\ddagger}}$ and continue to increase their profit until the price cap is reached, i.e., $\partial\tilde{C}_1^i\left(q_1^i\right)/\partial q_1^i\Big|_{q_1^i=q_1^{i\#}} = \overline{P}_1$, such that $q_1^{i\#} > q_1^{i\ddagger}$. Thus, in this case, market power is not completely eliminated. In Fig. 4, we illustrate the case above with the price cap $\overline{P}_1$, the marginal total declared cost $\partial\tilde{C}_1/\partial q_1$ and the oligopolistic $q^{\#}_1$ under the price cap.

### 4.2. Proposed incentive mechanism

#### 4.2.1. Design

In the previous section, we showed that price caps do not sufficiently remedy market power and do not address the problem of pollution damage. In order to effectively overcome both these problems jointly in the pricing mechanism, we propose an *incentive compatible* penalty and reward scheme for all producers. The proposed mechanism relies on providing each producer $i$ with an amount $\phi^i\left(\left[q^i_{nj}\right]_{nj}, \left[q^{i'}_{nj}\right]^{i'\neq i}_{nj}\right)$ as an incentive. We proceed to derive function $\phi^i\left(\left[q^i_{nj}\right]_{nj}, \left[q^{i'}_{nj}\right]^{i'\neq i}_{nj}\right)$ by requiring that, for every node $n$ and unit $j$, the profit maximising generation levels match the optimal generation levels $q^{i*}_{nj}$, i.e.,

$$\left[q^{i*}_{nj}\right]_{nj} \in \arg\max_{\left[q^i_{nj}\right]_{nj}} \left( \sum_{n'} \left( P_{n'}\left(\left[q_{n''}\right]_{n''}\right) q^i_{n'} - \sum_{j'} C^i_{n'j'}\left(q^i_{n'j'}\right) \right) + \phi^i\left(\left[q^i_{n'j'}\right]_{n'j'}, \left[q^{i'}_{n'j'}\right]^{i'\neq i}_{n'j'}\right) \right) \Bigg|_{\left[q^{i'}_{n'j'}\right]^{i'\neq i}_{n'j'} = \left[q^{i'*}_{n'j'}\right]^{i'\neq i}_{n'j'}} \tag{34}$$

subject to (1). Thus, $\phi^i\left(\left[q^i_{nj}\right]_{nj}, \left[q^{i'}_{nj}\right]^{i'\neq i}_{nj}\right)$ should be such that

$$\left( \sum_{n'} \frac{\partial P_{n'}\left(\left[q_{n''}\right]_{n''}\right)}{\partial q_n} q^i_{n'} + P_n\left(\left[q_{n'}\right]_{n'}\right) - \frac{\partial C^i_{nj}\left(q^i_{nj}\right)}{\partial q^i_{nj}} + \frac{\partial \phi^i\left(\left[q^i_{n'j'}\right]_{n'j'}, \left[q^{i'}_{n'j'}\right]^{i'\neq i}_{n'j'}\right)}{\partial q^i_{nj}} \right) \Bigg|_{\substack{q^i_{nj}=0,\ \left[q^i_{nj'}\right]_{j'\neq j} = \left[q^{i*}_{nj'}\right]_{j'\neq j}, \\ \left[q^{i'}_{nj'}\right]^{i'\neq i}_{j'} = \left[q^{i'*}_{nj'}\right]^{i'\neq i}_{j'}, \\ \left[q^{i'}_{n'j'}\right]^{i'}_{n'\neq nj'} = \left[q^{i'*}_{n'j'}\right]^{i'}_{n'\neq nj'}}} \le 0 \text{ if } q^{i*}_{nj} = 0,$$

$$\left( \sum_{n'} \frac{\partial P_{n'}\left(\left[q_{n''}\right]_{n''}\right)}{\partial q_n} q^i_{n'} + P_n\left(\left[q_{n'}\right]_{n'}\right) - \frac{\partial C^i_{nj}\left(q^i_{nj}\right)}{\partial q^i_{nj}} + \frac{\partial \phi^i\left(\left[q^i_{n'j'}\right]_{n'j'}, \left[q^{i'}_{n'j'}\right]^{i'\neq i}_{n'j'}\right)}{\partial q^i_{nj}} \right) \Bigg|_{\substack{q^i_{nj}=k^i_{nj},\ \left[q^i_{nj'}\right]_{j'\neq j} = \left[q^{i*}_{nj'}\right]_{j'\neq j}, \\ \left[q^{i'}_{nj'}\right]^{i'\neq i}_{j'} = \left[q^{i'*}_{nj'}\right]^{i'\neq i}_{j'}, \\ \left[q^{i'}_{n'j'}\right]^{i'}_{n'\neq nj'} = \left[q^{i'*}_{n'j'}\right]^{i'}_{n'\neq nj'}}} \ge 0 \text{ if } q^{i*}_{nj} = k^i_{nj},$$

$$\left( \sum_{n'} \frac{\partial P_{n'}\left(\left[q_{n''}\right]_{n''}\right)}{\partial q_n} q^i_{n'} + P_n\left(\left[q_{n'}\right]_{n'}\right) - \frac{\partial C^i_{nj}\left(q^i_{nj}\right)}{\partial q^i_{nj}} + \frac{\partial \phi^i\left(\left[q^i_{n'j'}\right]_{n'j'}, \left[q^{i'}_{n'j'}\right]^{i'\neq i}_{n'j'}\right)}{\partial q^i_{nj}} \right) \Bigg|_{\left[q^{i'}_{n'j'}\right]^{i'}_{n'j'} = \left[q^{i'*}_{n'j'}\right]^{i'}_{n'j'}} = 0 \text{ otherwise.} \tag{35}$$

By contrasting the above expressions with the characterisation of the optimal generation levels in (10), we obtain

$$\frac{\partial \phi^i\left(\left[q^i_{n'j'}\right]_{n'j'}, \left[q^{i'}_{n'j'}\right]^{i'\neq i}_{n'j'}\right)}{\partial q^i_{nj}} = \frac{\partial U\left(\left[q_{n'}\right]_{n'}\right)}{\partial q_n} - \sum_{n'} \frac{\partial P_{n'}\left(\left[q_{n''}\right]_{n''}\right)}{\partial q_n} q^i_{n'} - P_n\left(\left[q_{n'}\right]_{n'}\right) - \frac{\partial E_n(x_n)}{\partial x_n}\frac{\partial x^i_{nj}}{\partial q^i_{nj}}. \tag{36}$$

We can also refer to Table 1 and observe that the above expression contains the terms that differ between the optimal and oligopolistic conditions. Note that from (14) the price and marginal utility are equal, but have been represented separately for ease of integration in what follows. By construction, $\phi^i\left(\left[q^i_{n'j'}\right]_{n'j'}, \left[q^{i'}_{n'j'}\right]^{i'\neq i}_{n'j'}\right)$ is differentiable and concave in $q^i_{nj}$ for every $n$ and $j$. In order to obtain $\phi^i\left(\left[q^i_{n'j'}\right]_{n'j'}, \left[q^{i'}_{n'j'}\right]^{i'\neq i}_{n'j'}\right)$, we integrate the above starting with the case $n = 1$ and $j = 1$. Upon integration with respect to $q^i_{11}$, from (14) and recalling that $x^i_{nj}(0) = 0\ \forall n \in \mathcal{N}, i \in \mathcal{I}, j \in \mathcal{J}$, we obtain

$$\begin{aligned}
&\phi^i\left(q^i_{11}, \left[q^i_{1j'}\right]_{j'>1}, \left[q^i_{n'j'}\right]_{n'>1j'}, \left[q^{i'}_{n'j'}\right]^{i'\neq i}_{n'j'}\right) \\
&\quad -\phi^i\left(0, \left[q^i_{1j'}\right]_{j'>1}, \left[q^i_{n'j'}\right]_{n'>1j'}, \left[q^{i'}_{n'j'}\right]^{i'\neq i}_{n'j'}\right) \\
&= U\left(\left[q_{n'}\right]_{n'}\right) - U\left(q_1 - q^i_{11}, \left[q_{n'}\right]_{n'>1}\right) \\
&\quad - \sum_{n'} \left(P_{n'}\left(\left[q_{n''}\right]_{n''}\right) - P_{n'}\left(q_1 - q^i_{11}, \left[q_{n''}\right]_{n''>1}\right)\right) q^i_{n'} \\
&\quad - P_1\left(q_1 - q^i_{11}, \left[q_{n'}\right]_{n'>1}\right) q^i_{11} - E_1(x_1) + E_1\left(x_1 - x^i_{11}\right).
\end{aligned} \tag{37}$$

We then iterate over $j$ to obtain

$$\begin{aligned}
&\phi^i\left(\left[q^i_{1j'}\right]_{j'}, \left[q^i_{n'j'}\right]_{n'>1j'}, \left[q^{i'}_{n'j'}\right]^{i'\neq i}_{n'j'}\right) - \phi^i\left(0, \left[q^i_{n'j'}\right]_{n'>1j'}, \left[q^{i'}_{n'j'}\right]^{i'\neq i}_{n'j'}\right) \\
&= U\left(\left[q_{n'}\right]_{n'}\right) - U\left(q_1 - q^i_1, \left[q_{n'}\right]_{n'>1}\right) \\
&- \sum_{n'} \left(P_{n'}\left(\left[q_{n''}\right]_{n''}\right) - P_{n'}\left(q_1 - q^i_1, \left[q_{n''}\right]_{n''>1}\right)\right) q^i_{n'} \\
&- P_1\left(q_1 - q^i_1, \left[q_{n'}\right]_{n'>1}\right) q^i_1 - E_1(x_1) + E_1\left(x_1 - x^i_1\right).
\end{aligned} \tag{38}$$

Finally, we iterate over $n$ to obtain

$$\begin{aligned}
&\phi^i\left(\left[q^i_{n'j'}\right]_{n'j'}, \left[q^{i'}_{n'j'}\right]^{i'\neq i}_{n'j'}\right) = U\left(\left[q_{n'}\right]_{n'}\right) - U\left(\left[q_{n'} - q^i_{n'}\right]_{n'}\right) \\
&- \sum_{n'} \left(P_{n'}\left(\left[q_{n''}\right]_{n''}\right) q^i_{n'} - E_{n'}(x_{n'}) + E_{n'}\left(x_{n'} - x^i_{n'}\right)\right) + \phi^i\left(0, \left[q^{i'}_{n'j'}\right]^{i'\neq i}_{n'j'}\right).
\end{aligned} \tag{39}$$

where $\phi^i\left(0, \left[q^{i'}_{n'j'}\right]^{i'\neq i}_{n'j'}\right)$ is a fixed amount by which $\phi^i\left(\left[q^i_{n'j'}\right]_{n'j'}, \left[q^{i'}_{n'j'}\right]^{i'\neq i}_{n'j'}\right)$ can be offset.

Effectively, the amount $\phi^i\left(\left[q^i_{n'j'}\right]_{n'j'}, \left[q^{i'}_{n'j'}\right]^{i'\neq i}_{n'j'}\right) - \phi^i\left(0, \left[q^{i'}_{n'j'}\right]^{i'\neq i}_{n'j'}\right)$ is equivalent to a subsidy equal to the total marginal consumer utility created at all nodes by $\left[q^i_n\right]_n$ minus the price, providing an incentive to producers to maximise utility, and a tax equal to the marginal externality due to $\left[x^i_n\right]_n$, transferring this cost to the producers at all nodes. If $\phi^i\left(\left[q^i_{n'j'}\right]_{n'j'}, \left[q^{i'}_{n'j'}\right]^{i'\neq i}_{n'j'}\right) > \phi^i\left(0, \left[q^{i'}_{n'j'}\right]^{i'\neq i}_{n'j'}\right)$, it is a net subsidy and incentivises producer $i$ to increase their generation levels compared to the oligopolistic case and if $\phi^i\left(\left[q^i_{n'j'}\right]_{n'j'}, \left[q^{i'}_{n'j'}\right]^{i'\neq i}_{n'j'}\right) < \phi^i\left(0, \left[q^{i'}_{n'j'}\right]^{i'\neq i}_{n'j'}\right)$ it is a net tax and incentivises producer $i$ to decrease their generation levels.

We can characterise the declared cost function of producer $i$ under the proposed mechanism with the aim of obtaining the optimal generation levels from (10) by reconsidering the spot market generation levels in (13), and obtain

$$\frac{\partial \tilde{C}^i_n\left(q^i_n\right)}{\partial q^i_n} = \frac{\partial C^i_{nj}\left(q^i_{nj}\right)}{\partial q^i_{nj}} + \frac{\partial E_n(x_n)}{\partial x_n}\frac{\partial x^i_{nj}}{\partial q^i_{nj}} \ge \frac{\partial C^i_{nj}\left(q^i_{nj}\right)}{\partial q^i_{nj}}. \tag{40}$$

Observe that $\tilde{C}^i_n\left(q^i_n\right)$ is by construction continuous, piecewise differentiable, increasing and convex in $q^i_n$. We can obtain an analytic expression for $\tilde{C}^i_n\left(q^i_n\right)$ by recalling that $C^i_{nj}(0) = 0\ \forall n \in \mathcal{N}, i \in \mathcal{I}, j \in \mathcal{J}$, enforcing $\tilde{C}^i_n(0) = 0\ \forall n \in \mathcal{N}, i \in \mathcal{I}$, and by integration with respect to $q^i_n$,

$$\tilde{C}^i_n\left(q^i_n\right) = \min_{\left[\overline{q}^i_{nj}\right]_j} \left( \sum_{j'} C^i_{nj'}\left(\overline{q}^i_{nj'}\right) + E_n\left(x_n\left(\left[q^{i'}_{nj'}\right]^{i'<i}_{j'}, \left[\overline{q}^i_{nj'}\right]_{j'}, \left[q^{i'}_{nj'}\right]^{i'>i}_{j'}\right)\right)\right.$$

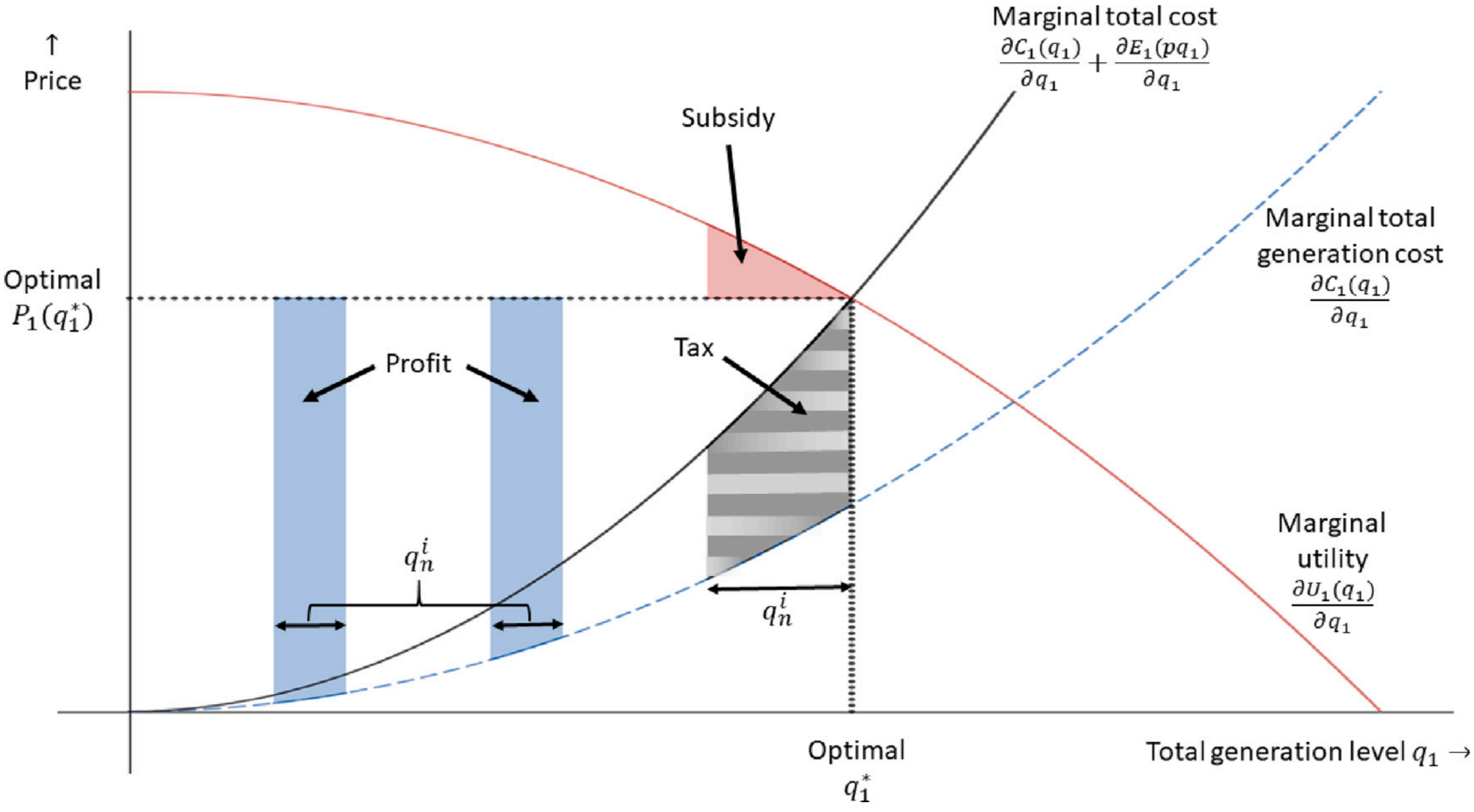

**Fig. 5.** Proposed incentive mechanism.

$$- E_n \left( x_n \left( \left[ q_{nj'}^{i'} \right]_{j'}^{i'<i}, \left[ \overline{q}_{nj'}^{i} \right]_{j'}, \left[ q_{nj'}^{i'} \right]_{j'}^{i'>i} \right) - x_n^i \left( \left[ \overline{q}_{nj'}^{i} \right]_{j'} \right) \right) \tag{41}$$

$$\text{subject to } \overline{q}_n^i = q_n^i \tag{42}$$

and the generator capacity constraint (1) which includes the externality of pollution. It is easy to verify that if the ISO solves (13) based on the declared cost function (41) then it solves (10).

For illustration, consider the example developed in Section 3. In Fig. 5, we take the example of a producer $i$ with generation level $q_1^i$. The solid shaded regions in the figure correspond to producer $i$'s profit from (25), the dotted region corresponds to the reward from (39), and the striped region corresponds to the penalty. Observe that the SaT are allotted based on the marginal generation level.

### 4.2.2. Properties

In this section, we discuss the properties, advantages and disadvantages of the proposed mechanism. First, $\phi^i \left( \left[ q_{nj}^i \right]_{nj}, \left[ q_{nj}^{i'} \right]_{nj}^{i' \neq i} \right)$ can be computed based on quantities $\left[ q_{n'}^{i'} \right]_{n'}^{i'}$ and $\left[ x_{n'}^{i'} \right]_{n'}^{i'}$ that the ISO can observe if and only if, for every producer $i$, $\phi^i \left( 0, \left[ q_{nj}^{i'} \right]_{nj}^{i' \neq i} \right)$ is chosen to depend only on the total quantities $\left[ q_{n'}^{i'} \right]_{n'}^{i' \neq i}$ and $\left[ x_{n'}^{i'} \right]_{n'}^{i' \neq i}$. If so, from (39) we can see that $\phi^i \left( \left[ q_{nj}^i \right]_{nj}, \left[ q_{nj}^{i'} \right]_{nj}^{i' \neq i} \right)$ depends only on the total quantities $\left[ q_{n'}^{i} \right]_{n'}$, $\left[ x_{n'}^{i} \right]_{n'}$, $\left[ q_{n'}^{i'} \right]_{n'}^{i' \neq i}$ and $\left[ x_{n'}^{i'} \right]_{n'}^{i' \neq i}$ and information asymmetry does not affect the mechanism.

Second, the mechanism is *non-discriminatory*, i.e., we have $\phi^i \left( \left[ q_{nj}^i \right]_{nj}, \left[ q_{nj}^{i'} \right]_{nj}^{i' \neq i} \right) \equiv \phi \left( \left[ q_{nj}^i \right]_{nj}, \left[ q_{nj}^{i'} \right]_{nj}^{i' \neq i} \right)$ if and only if $\phi^i \left( 0, \left[ q_{nj}^{i'} \right]_{nj}^{i' \neq i} \right) \equiv \phi \left( 0, \left[ q_{nj}^{i'} \right]_{nj}^{i' \neq i} \right)$ for every producer $i$, i.e., for every producer $i$, we choose $\phi^i \left( 0, \left[ q_{nj}^{i'} \right]_{nj}^{i' \neq i} \right)$ that is not parametrised by $i$. The result follows directly from (39). Note that $\phi^i \left( \left[ q_{nj}^i \right]_{nj}, \left[ q_{nj}^{i'} \right]_{nj}^{i' \neq i} \right)$ would retain its dependence on producers' output $\left[ q_{nj}^i \right]_{nj}$ and $\left[ q_{nj}^{i'} \right]_{nj}^{i' \neq i}$, but not on the producer $i$.

Third, for every producer $i$, participation in the market is *individually rational*, i.e., if producer $i$'s participation does not decrease social welfare then producer $i$ achieves a net profit, if and only if

$$\begin{aligned} \phi^i \left( 0, \left[ q_{n'j'}^{i'*} \right]_{n'j'}^{i' \neq i} \right) &\geq -U \left( \left[ q_{n'}^* \right]_{n'} \right) + U \left( \left[ q_{n'}^* - q_{n'}^{i*} \right]_{n'} \right) \\ &+ \sum_{n'} \left( C_{n'}^i \left( q_{n'}^{i*} \right) + E_{n'} \left( x_{n'}^* \right) - E_{n'} \left( x_{n'}^* - x_{n'}^{i*} \right) \right) \end{aligned} \tag{43}$$

where $x_{nj}^{i*} = x_{nj}^i \left( q_{nj}^{i*} \right) \; \forall n \in \mathcal{N}, i \in \mathcal{I}, j \in \mathcal{J}$. The result follows directly from (39). Given that the optimal generation levels $\left[ q_{nj}^{i*} \right]_{nj}^{i}$ obey

$$\begin{aligned} & U \left( \left[ q_{n'}^* \right]_{n'} \right) - U \left( \left[ q_{n'}^* - q_{n'}^{i*} \right]_{n'} \right) \\ & - \sum_{n'} \left( C_{n'}^i \left( q_{n'}^{i*} \right) + E_{n'} (x_{n'}^*) - E_{n'} \left( x_{n'}^* - x_{n'}^{i*} \right) \right) \geq 0 \end{aligned} \tag{44}$$

making the choice $\phi^i \left( 0, \left[ \overline{q}_{n'j'}^{i'} \right]_{n'j'}^{i' \neq i} \right) \equiv \phi \left( 0, \left[ q_{n'}^{i'} \right]_{n'}^{i' \neq i} \right)$ such that $\sum_j \overline{q}_{nj}^i = q_n^i \; \forall n \in \mathcal{N}, i \in \mathcal{I}$, and $\phi \left( 0, \left[ q_{n'}^{i'} \right]_{n'}^{i' \neq i} \right) \geq 0$ satisfies all the conditions presented above on $\phi^i \left( 0, \left[ q_{n'j'}^{i'} \right]_{n'j'}^{i' \neq i} \right)$.

Fourth, the mechanism may create funding problems for the ISO if it has to grant a net subsidy, i.e., $\sum_i \phi^i \left( \left[ q_{nj}^i \right]_{nj}, \left[ q_{nj}^{i'} \right]_{nj}^{i' \neq i} \right) \geq 0$. This may also result in potentially large profits for producers. In practice, funding problems are overcome by charging fixed fees to producers and other market participants, viz., consumers and merchants. The greater the fixed fee, the lower the funding problems. Observe that, for producer $i$, the quantity $-\phi^i \left( 0, \left[ q_{n'j'}^{i'} \right]_{n'j'}^{i' \neq i} \right)$ can be regarded as a fixed fee. However, from (43), setting a high value of $-\phi^i \left( 0, \left[ q_{n'j'}^{i'} \right]_{n'j'}^{i' \neq i} \right)$ would come at a cost to individual rationality.

Finally, the proposed mechanism, if used simultaneously with price caps, eliminates the disadvantages of price caps. This is because producer $i$ maximises the marginal social welfare created by $\left[ q_{nj}^i \right]_{nj}$ which is independent of the price $P_n \left( \left[ q_{n'} \right]_{n'} \right)$ and hence is not affected by the price cap at node $n$.

## 5. Analytical and numerical example

Consider a two-node, two-producer, two-unit example where $\mathcal{N} = \{1,2\}$, $\mathcal{G} = \{1,2\}$ and $\mathcal{J} = \{1,2\}$. The nodes are connected by a single transmission line where $\mathcal{L} = \{1 \leftrightarrow 2\}$, and $f_{1\leftrightarrow 2} = 5$. The parameters and functions for every $n$, $i$ and $j$ are as follows.

| $(n,i,j)$ | (1,1,1) | (1,1,2) | (1,2,1) | (1,2,2) | (2,1,1) | (2,1,2) | (2,2,1) | (2,2,2) |
|---|---|---|---|---|---|---|---|---|
| $k^i_{nj}$ | 5 | 5 | 10 | 10 | 5 | 5 | 10 | 10 |
| $C^i_{nj}\left(q^i_{nj}\right)$ | $2q^1_{11}$ | $q^1_{12}$ | $2q^2_{11}$ | $q^2_{12}$ | $4q^1_{21}$ | $2q^1_{22}$ | $4q^2_{21}$ | $2q^2_{22}$ |
| $x^i_{nj}\left(q^i_{nj}\right)$ | $q^1_{11}$ | $3q^1_{12}$ | $q^2_{11}$ | $3q^2_{12}$ | $q^1_{21}$ | $3q^1_{22}$ | $q^2_{21}$ | $3q^2_{22}$ |

| $n$ | $E_n(x_n)$ | $\tilde{U}_n(d_n)$ | $H_{1\leftrightarrow 2n}$ |
|---|---|---|---|
| 1 | $x_1$ | $-(d_1)^2 + 44d_1$ if $d_1 \leq 22$, $22^2$ otherwise | 1 |
| 2 | $2x_2$ | $6d_2$ | 0 |

We can define the optimal power consumption as

$$\left[d^*_n\right]_n \left(\left[q_{n'}\right]_{n'}\right) = \arg\max_{[d_n]_n} \sum_{n'} \tilde{U}_{n'}(d_{n'}) \tag{45}$$

subject to the constraints (2), (4) and (5). Based on the above, we can determine the optimal demand,

$$\left(d^*_1(q_1,q_2), d^*_2(q_1,q_2)\right) = \left(\max(\min(q_1+\min(q_2,5),19), q_1-5), \max(0, q_2-5, q_2+\min(q_1-19,5))\right) \tag{46}$$

and accordingly the utility function from (7),

$$U(q_1,q_2) = \begin{cases} -(q_1+q_2)^2 + 44(q_1+q_2) & \text{if } q_1+q_2 \leq 19, q_2 \leq 5, \\ -(q_1+5)^2 + 44(q_1+5) + 6(q_2-5) & \text{if } q_1 \leq 14, q_2 > 5, \\ -19^2 + 44\times 19 + 6(q_1+q_2-19) & \text{if } q_1+q_2 > 19, 14 < q_1 \leq 24, \\ -(q_1-5)^2 + 44(q_1-5) + 6(q_2+5) & \text{if } 24 < q_1 \leq 27, \\ 22^2 + 6(q_2+5) & \text{otherwise}, \end{cases} \tag{47}$$

the price at each $n$ from (14),

$$P_1\left(q_1,q_2\right) = \begin{cases} -2(q_1+q_2)+44 & \text{if } q_1+q_2 \leq 19, q_2 \leq 5, \\ -2(q_1+5)+44 & \text{if } q_1 \leq 14, q_2 > 5, \\ 6 & \text{if } q_1+q_2 > 19, 14 < q_1 \leq 24, \\ -2(q_1-5)+44 & \text{if } 24 < q_1 \leq 27, \\ 0 & \text{otherwise}, \end{cases} \tag{48}$$

$$P_2\left(q_1,q_2\right) = \begin{cases} -2(q_1+q_2)+44 & \text{if } q_1+q_2 \leq 19, q_2 \leq 5, \\ 6 & \text{otherwise}, \end{cases} \tag{49}$$

each producer $i$'s total cost function at each $n$ from (16),

$$C^1_1\left(q^1_1\right) = q^1_1\left(1-\Theta\left(q^1_1-5\right)\right) + \left(2q^1_1-5\right)\Theta\left(q^1_1-5\right), \tag{50}$$

$$C^2_1\left(q^2_1\right) = q^2_1\left(1-\Theta\left(q^2_1-10\right)\right) + \left(2q^2_2-10\right)\Theta\left(q^2_1-10\right), \tag{51}$$

$$C^1_2\left(q^1_2\right) = 2q^1_2\left(1-\Theta\left(q^1_2-5\right)\right) + \left(4q^1_2-10\right)\Theta\left(q^1_2-5\right), \tag{52}$$

$$C^2_2\left(q^2_2\right) = 2q^2_2\left(1-\Theta\left(q^2_2-10\right)\right) + \left(4q^2_2-20\right)\Theta\left(q^2_2-10\right) \tag{53}$$

where $\Theta(x)$ is the Heaviside step function,[5] and each producer $i$'s declared cost function at each $n$ under market power from (28)

$$\tilde{C}^1_1\left(q^1_1\right) = C^1_1\left(q^1_1\right) - P_1(q_1,q_2)q^1_1 - P_2(q_1,q_2)q^1_2 + P_2\left(q^2_1,q_2\right)q^1_2 + U(q_1,q_2) - U(q^2_1,q_2), \tag{54}$$

$$\tilde{C}^2_1\left(q^2_1\right) = C^2_1\left(q^2_1\right) - P_1(q_1,q_2)q^2_1 - P_2(q_1,q_2)q^2_2 + P_2\left(q^1_1,q_2\right)q^2_2 + U(q_1,q_2) - U(q^1_1,q_2), \tag{55}$$

$$\tilde{C}^1_2\left(q^1_2\right) = C^1_2\left(q^1_2\right) - P_1(q_1,q_2)q^1_1 - P_2(q_1,q_2)q^1_2 + P_1\left(q_1,q^2_2\right)q^1_1 + U(q_1,q_2) - U(q_1,q^2_2), \tag{56}$$

$$\tilde{C}^2_2\left(q^2_2\right) = C^2_2\left(q^2_2\right) - P_1(q_1,q_2)q^2_1 - P_2(q_1,q_2)q^2_2 + P_1\left(q_1,q^1_2\right)q^2_1 + U(q_1,q_2) - U(q_1,q^1_2). \tag{57}$$

[5] The Heaviside step function is defined as $\Theta(x) = \begin{cases} 0 & \text{if } x < 0, \\ 1 & \text{otherwise.} \end{cases}$

**Table 2**
Generation levels, social welfare and prices under different solutions.

| Generation for $(n,i,j)$ | Optimal $q^{i*}_{nj}$ | Competitive $q^{i\ddagger}_{nj}$ | Oligopolistic $q^{i\#}_{nj}$ | Price cap $\overline{P}_n = 8$ $q^{i\#}_{nj}$ | Price cap $\overline{P}_n = 1$ $q^{i\#}_{nj}$ |
|---|---|---|---|---|---|
| (1,1,1) | 5 | 5 | 0 | 0 | 0 |
| (1,1,2) | 5 | 5 | 2.76 | 5 | 5 |
| (1,2,1) | 10 | 6 | 0 | 0 | 0 |
| (1,2,2) | 5 | 10 | 2.76 | 8 | 10 |
| (2,1,1) | 5 | 5 | 0 | 0 | 0 |
| (2,1,2) | 0 | 5 | 2.37 | 2.5 | 0 |
| (2,2,1) | 10 | 10 | 0 | 0 | 0 |
| (2,2,2) | 0 | 10 | 2.37 | 2.5 | 0 |
| Social welfare | 425 | 390 | 286.1724 | 376 | 375 |
| Price $P_1$ | 4 | 2 | 23.48 | 8 | 1 |
| Price $P_2$ | 6 | 6 | 23.48 | 8 | 1 |

Accordingly, we obtain the optimal, competitive and oligopolistic generation levels and the corresponding social welfare in each case as stated in Table 2. Here, the *optimal prices* are obtained from (15) under the assumption that the cost function is declared such that optimal solution is achieved. We will determine this declared cost function later in this section.

To see how we obtained these results, consider (32) which generalises the conditions to obtain the optimal, competitive and generation levels. The generation levels $\left[q^{i\Box}_{nj}\right]^i_{nj}$ can be obtained by simultaneously solving

$$f^i_{nj}\left(\left[q^{i'\Box}_{n',j'}\right]^{i'}_{n'j'}\right) = 0 \; \forall n, \forall i, \forall j. \tag{58}$$

If the result is $q^{i\Box}_{nj} < 0$, the generation capacity constraint (1) is violated. As the problems are convex, the solution must be $q^{i\Box}_{nj} = 0$. Similarly, if the result is $q^{i\Box}_{nj} > k^{i\Box}_{nj}$, the solution must be $q^{i\Box}_{nj} = k^{i\Box}_{nj}$. The corrected solutions can be used to solve (58) iteratively until the results converge. Recall that the terms that are represented by $f^i_{nj}\left(\left[q^{i'}_{n',j'}\right]^{i'}_{n'j'}\right)$ would depend upon which solution we wish to obtain and are represented in Table 1, e.g. to obtain the oligopolistic generation levels respectively, solving (58) for bus $n$ and producer $i$ is equivalent to obtaining the intersection between the price at bus $n$ and the marginal declared cost function at bus $n$ of producer $i$. Given this, the generation levels for each individual technology $j$ can be obtained from (18).

As can be seen, the social welfare is maximal in the optimal case. In the competitive case, the generation levels are higher than in the optimal case, but the social welfare and prices are lower. This is because the competitive market clearing does not consider the pollution damage and the associated loss in social welfare. Also, as the costs considered are fewer, the prices are lower. In the oligopolistic case, the generation levels are lesser and accordingly, the prices are higher than in the optimal case showing that market power is not always effective in alleviating the externality due to pollution. Also, observe that, in the optimal case, the generation levels are determined on the basis of total cost (including the negative externality) where unit 1 is utilised to its maximum capacity as opposed to the only generation cost in the other cases where unit 2 is preferred.

In what follows, we compare price caps to our proposed tax and subsidy scheme. First, consider price caps as a potential solution. Consider that the ISO has set the price caps $\overline{P_1} = \overline{P_2} = 8$. As the price cap is lesser than the oligopolistic prices and higher than the competitive prices at both busses, we can assume that the marginal utility curve would intersect piece 2 of the marginal declared cost curve which is listed in Section 4.1. Accordingly, we obtain the generation levels as stated in Table 2.

Now, consider that the ISO has set the price caps $\overline{P}_1 = \overline{P}_2 = 1$. As the price cap is not greater than the competitive price at both busses, the marginal utility curve would intersect piece 3 of the marginal declared cost curve. However, as the intersection would lie above the price cap, the generation levels will be $\hat{q}^i_n$ (determined by the intersection of the price cap with the marginal true cost functions) as in Fig. 3 and as stated in Table 2. The consumers' desired demand at each bus $n$ will be determined by the intersection of the price cap with the marginal utility curves of the bus $n$, i.e., at $d^{\S}_n$ such that $\partial \tilde{U}_n / \partial d_n |_{d_n = d^{\S}_n} = \overline{P}_n$. For bus 1, the desired demand is $d^{\S}_1 = 26.5$ and for bus 2, it is $d^{\S}_2 \to \infty$ as the marginal utility is $\partial \tilde{U}_2 / \partial d_2 = 6$ for all values of $d_2$ which is greater than the price cap. Therefore, there will be involuntary load shedding at both busses. So far we have considered examples of both, high and low price caps respectively. In both cases, we observe that the social welfare is less than the optimal case. Also, the problem of pollution damage is not addressed.

Next, we consider our proposed tax and subsidy scheme. The amount provided to each producer $i$ as an incentive from (39),

$$\phi^1\left(\left[q^1_{nj}\right]_{nj}, \left[q^2_{nj}\right]_{nj}\right) = U(q_1, q_2) - U\left(q^2_1, q^2_2\right) - P_1(q_1, q_2)q^1_1 - P_2(q_1, q_2)q^1_2 - q^1_{11} - 3q^1_{12} - 2q^1_{21} - 6q^1_{22} + \phi^1\left(0, \left[q^2_{nj}\right]_{nj}\right), \quad (59)$$

$$\phi^2\left(\left[q^2_{nj}\right]_{nj}, \left[q^1_{nj}\right]_{nj}\right) = U(q_1, q_2) - U\left(q^1_1, q^1_2\right) - P_1(q_1, q_2)q^2_1 - P_2(q_1, q_2)q^2_2 - q^2_{11} - 3q^2_{12} - 2q^2_{21} - 6q^2_{22} + \phi^2\left(0, \left[q^1_{nj}\right]_{nj}\right), \quad (60)$$

and each producer $i$'s declared cost function at each $n$ under the proposed mechanism from (41),

$$\tilde{C}^1_1\left(q^1_1\right) = 3q^1_1\left(1 - \Theta\left(q^1_1 - 5\right)\right) + \left(4q^1_1 - 5\right)\Theta\left(q^1_1 - 5\right), \quad (61)$$

$$\tilde{C}^2_1\left(q^2_1\right) = 3q^2_1\left(1 - \Theta\left(q^2_1 - 10\right)\right) + \left(4q^2_1 - 10\right)\Theta\left(q^2_1 - 10\right), \quad (62)$$

$$\tilde{C}^1_2\left(q^1_2\right) = 6q^1_2\left(1 - \Theta\left(q^1_2 - 5\right)\right) + \left(8q^1_2 - 10\right)\Theta\left(q^1_2 - 5\right), \quad (63)$$

$$\tilde{C}^2_2\left(q^2_2\right) = 6q^2_2\left(1 - \Theta\left(q^2_2 - 10\right)\right) + \left(8q^2_2 - 20\right)\Theta\left(q^2_2 - 10\right). \quad (64)$$

The declared cost function above would result in the optimal solution and can be used to determine the optimal prices from (15).

The amount of tax and subsidy charged will be based on the optimal generation levels as that would be the outcome. For producers 1 and 2, respectively, the amount is

$$\phi^1\left(\left[q^{1*}_{nj}\right]_{nj}, \left[q^{2*}_{nj}\right]_{nj}\right) = -11 + \phi^1\left(0, \left[q^{2*}_{nj}\right]_{nj}\right), \quad (65)$$

$$\phi^2\left(\left[q^{1*}_{nj}\right]_{nj}, \left[q^{2*}_{nj}\right]_{nj}\right) = 0 + \phi^2\left(0, \left[q^{1*}_{nj}\right]_{nj}\right). \quad (66)$$

As we have shown in the previous section, unlike price caps, our proposed tax and subsidy scheme effectively ensures that the pollution damage is addressed and mitigates market power resulting in the optimal solution.

The example we have provided above is for a simple system to illustrate the theory as developed. Whilst it is out of scope to undertake a full case study with real data, we would consider it to be quite feasible in future work. For example, the market data including the power grid parameters, consumer utilities, producer costs and generation levels for a few European countries, e.g., Sweden and Germany can be obtained from Nordpool (2023). European Environment Agency (2023) contains data about pollution from combustion plants all over Europe. Comparing these pollution levels to the plants' generation levels, we may estimate pollution as a function of generation. Pollution damage as a function of pollution can be estimated from Watkiss et al. (2005).

## 6. Conclusion

In this research, we have modelled an electricity market clearing process which seeks to maximise the social welfare including the producer and consumer utilities as well as the negative externality due to pollution. We have identified two problems associated with achieving this social optimal solution. The first is that due to the ISO's incomplete information, electricity markets essentially create a competitive market clearing process which does not include pollution externalities. This results in competitive generation levels that are greater than the optimal levels. The second is producers' market power leading to imperfect competition which results in oligopolistic generation levels that differ from the optimal levels. Price caps are commonly used to overcome the latter, but we have shown that they are insufficient when they are set too high and can create inconsistencies in the market if they are set too low. We have proposed to tax and subsidise producers by their marginal contribution to the pollution damage and consumer surplus respectively to overcome both the problems respectively.

We have argued, in Section 3.2, that due to information asymmetry the pollution cannot be represented as a function of generation levels and, therefore, it cannot be included directly in the competitive market clearing price. Had it somehow been included in the price without our proposed tax, it would encourage producers to increase their pollution so that prices and hence their profits would increase. This is another reason why it is important to have a tax such as the one we have proposed to transfer the externality to the producers.

Our proposed tax and subsidy scheme can be implemented on the basis of quantities observable by the ISO and can therefore be implemented without requiring significant systematic changes. Also, as it is individually rational, we do not need to devise another means of encouraging participation. As it is compatible with price caps and even eliminates its disadvantages, markets with price caps can retain them. Finally, electricity spot markets are cleared by ISOs using large scale optimisation models and so the type of incentive formulation proposed here would be a relatively small adaptation to the already complex algorithms in use.

In the appendices, we show that the proposed incentive mechanism can be extended to overcome multiple externalities and even accommodate additional constraints such as temporal constraints due to energy limited technologies.

## CRediT authorship contribution statement

**Lamia Varawala:** Conceptualization, Methodology, Writing – original draft, Visualization. **Mohammad Reza Hesamzadeh:** Conceptualization, Methodology, Supervision, Project administration, Funding acquisition. **György Dán:** Methodology, Writing – review & editing, Supervision, Project administration, Funding acquisition. **Derek Bunn:** Conceptualization, Writing – review & editing. **Juan Rosellón:** Writing – review & editing.

## Declaration of competing interest

The authors declare that they have no known competing financial interests or personal relationships that could have appeared to influence the work reported in this paper.

## Appendix A. Multiple externalities

Throughout the article, we parametrised the generation level $q^i_{nj}$ by unit $j$ in order to be able to make the amount of pollution $x^i_{nj}\left(q^i_{nj}\right)$ dependent on the generation unit employed. We then attributed a negative externality $E_n(x_n)$ to pollution at every node $n$. We can extend the model to other properties of production facilities and the resulting externalities by introducing indices $m, m'$, and by using them to parametrise the generation level $q^i_{njm}$, such that $q^i_{nm} = \sum_j q^i_{njm}$, and the generation cost function $C^i_{njm}\left(q^i_{njm}\right)$. Then, for each node $n$, producer $i$ and property $m$, we define the quantity $y^i_{nm}\left(q^i_{nm}\right)$ associated to the property. Similar to $x^i_{nj}\left(q^i_{nj}\right)$, we assume that $y^i_{nm}\left(q^i_{nm}\right)$ is differentiable, increasing and convex in $q^i_{nm}$ and that the ISO can observe the total value $y^i_n$. We can redefine the total externality as $E\left(\left[x_n\right]_n, \left[y_n\right]_n\right)$ and assume that it is differentiable and convex in $x_n$ and $y_n$. It is easy to see that $E\left(\left[x_n\right]_n, \left[y_n\right]_n\right)$ is increasing in $y_n$ if and only if $y_n$ causes a negative externality. Following the analysis in this article, the proposed mechanism becomes

$$\begin{aligned}
&\phi^i\left(\left[q^i_{n'j'm'}\right]_{n'j'm'}, \left[q^{i'}_{n'j'm'}\right]^{i'\neq i}_{n'j'm'}\right) = U\left(\left[q_{n'}\right]_{n'}\right) - U\left(\left[q_{n'} - q^i_{n'}\right]_{n'}\right) \\
&\quad - \sum_{n'} P_{n'}\left(\left[q_{n''}\right]_{n''}\right) q^i_{n'} - E(\left[x_{n'}\right]_{n'}, \left[y_{n'}\right]_{n'}) + E\left(\left[x_{n'} - x^i_{n'}\right]_{n'}, \left[y_{n'} - y^i_{n'}\right]_{n'}\right) \\
&\quad + \phi^i\left(0, \left[q^{i'}_{n'j'm'}\right]^{i'\neq i}_{n'j'm'}\right).
\end{aligned} \tag{67}$$

## Appendix B. Convex constraints on the ISO

In (7), we defined the optimal utility subject to constraints (2), (4) and (5) on the ISO. Generation capacity constraint (1) was not included because the ISO may not be able to observe $\left[q^i_{nj}\right]^i_{nj}$ or $\left[k^i_{nj}\right]^i_{nj}$ due to information asymmetry. However, contrary to constraints (2), (4) and (5), producers explicitly account for constraint (1) in the oligopolistic generation levels in (24), and hence (1) will be satisfied too.

We can extend the model by including additional convex constraints on the ISO, akin to (2), (4) and (5), of the form

$$h\left(\left[q^i_n\right]^i_n, \left[d_n\right]_n\right) \leq 0. \tag{68}$$

As such a constraint is not explicitly imposed on the producers, we must redefine the optimal utility as (7) subject to (2), (4), (5), and (68) in order to implicitly account for it. Note that the SaT mechanism defined in (39) and the declared cost function in (41) remain the same.

## Appendix C. Energy-limited technologies and generalised additional constraints on producers

Technologies such as solar power plants and hydroelectric turbines may have an energy limit that imposes a constraint on the total energy produced over a certain duration of time on producers in addition to the generation capacity constraints that would apply to every time instant.

To model the additional constraint, we use $t, t' \in \mathbb{N}_0, 0 < t, t' \leq T$ for indexing the intervals in the spot market and all time-varying quantities $q^{it}_{nj}$, $x^{it}_{nj}\left(q^{it}_{nj}\right)$, $U^t\left(\left[q^t_n\right]_n\right)$, $P^t_n\left(\left[q^t_{n'}\right]_{n'}\right)$ and $\phi^{it}\left(\left[q^{it}_{nj}\right]_{nj}, \left[q^{i't}_{nj}\right]^{i'\neq i}_{nj}\right)$. For clarity, we will explicitly represent a summation over $t$ instead of omitting the index. Furthermore, for each node $n$, producer $i$ and unit $j$, we denote by $g^i_{nj} \geq 0$ the limit on energy produced over $T$ intervals of time normalised by the duration of a single interval, where

$$\sum_{t=1}^{T} q^{it}_{nj} \leq g^i_{nj}. \tag{69}$$

In addition, we denote by $\hat{C}^i_{nj}$ the generation cost function in this section for reasons that will become evident shortly.

Using these notations the optimal generation levels are given by

$$\begin{aligned}
&\left[q^{it*}_{nj}\right]^{it}_{nj} \in \arg\max_{\left[q^{it*}_{nj}\right]^{it}_{nj}} \sum_{t'=1}^{T}\left(U^{t'}\left(\left[q^{t'}_{n'}\right]^{t'}_{n'}\right)\right. \\
&\quad \left. - \sum_{n'}\left(\sum_{i'j'} \hat{C}^{i'}_{n'j'}\left(q^{i't'}_{n'j'}\right) + E_{n'}\left(x^{t'}_{n'}\right)\right)\right)
\end{aligned} \tag{70}$$

subject to

$$0 \leq q^{it}_{nj} \leq k^{it}_{nj} \tag{71}$$

and (69). Observe that because the ISO maximises social welfare, any constraint that applies to producers must also be considered by the ISO. We can characterise the optimal generation levels as

$$q^{it*}_{nj} \in \begin{cases}
\{0\} & \text{if } \left(\sum_{n'} \frac{\partial U^t\left(\left[q^t_{n'}\right]_{n'}\right)}{\partial q^t_n} - \frac{\partial \hat{C}^i_{nj}\left(q^{it}_{nj}\right)}{\partial q^{it}_{nj}} - \frac{\partial E_n(x^t_n)}{\partial x^t_n}\frac{\partial x^{it}_{nj}}{\partial q^{it}_{nj}} - \lambda^i_{nj}\right)\Bigg|_{\substack{q^{it}_{nj}=0, \left[q^{it'}_{nj}\right]^{t'} = \left[q^{it'\neq t*}_{nj}\right]^{t'}, \left[q^{it'}_{nj'}\right]^{t'}_{j'\neq j} = \left[q^{it'*}_{nj'}\right]^{t'}_{j'\neq j}, \\ \left[q^{i't'}_{nj'}\right]^{i'\neq it'}_{j'} = \left[q^{i't'*}_{nj'}\right]^{i'\neq it'}_{j'}, \left[q^{i't'}_{n'j'}\right]^{i't'}_{n'\neq nj'} = \left[q^{i't'*}_{n'j'}\right]^{i't'}_{n'\neq nj'}}} < 0, \\
\left\{k^i_{nj}\right\} & \text{if } \left(\sum_{n'} \frac{\partial U^t\left(\left[q^t_{n'}\right]_{n'}\right)}{\partial q^t_n} - \frac{\partial \hat{C}^i_{nj}\left(q^{it}_{nj}\right)}{\partial q^{it}_{nj}} - \frac{\partial E_n(x^t_n)}{\partial x^t_n}\frac{\partial x^{it}_{nj}}{\partial q^{it}_{nj}} - \lambda^i_{nj}\right)\Bigg|_{\substack{q^{it}_{nj}=k^{it}_{nj}, \left[q^{it'}_{nj}\right]^{t'} = \left[q^{it'\neq t*}_{nj}\right]^{t'}, \left[q^{it'}_{nj'}\right]^{t'}_{j'\neq j} = \left[q^{it'*}_{nj'}\right]^{t'}_{j'\neq j}, \\ \left[q^{i't'}_{nj'}\right]^{i'\neq it'}_{j'} = \left[q^{i't'*}_{nj'}\right]^{i'\neq it'}_{j'}, \left[q^{i't'}_{n'j'}\right]^{i't'}_{n'\neq nj'} = \left[q^{i't'*}_{n'j'}\right]^{i't'}_{n'\neq nj'}}} > 0, \\
\left[0, k^i_{nj}\right] & \text{s.t. } \left(\sum_{n'} \frac{\partial U^t\left(\left[q^t_{n'}\right]_{n'}\right)}{\partial q^t_n} - \frac{\partial \hat{C}^i_{nj}\left(q^{it}_{nj}\right)}{\partial q^{it}_{nj}} - \frac{\partial E_n(x^t_n)}{\partial x^t_n}\frac{\partial x^{it}_{nj}}{\partial q^{it}_{nj}} - \lambda^i_{nj}\right)\Bigg|_{\left[q^{i't'}_{n'j'}\right]^{i't'}_{n'j'} = \left[q^{i't'*}_{n'j'}\right]^{i't'}_{n'j'}} = 0 \text{ otherwise},
\end{cases} \tag{72}$$

where $\lambda^i_{nj}$ is the *Lagrange multiplier* associated to (69) for node $n$, producer $i$ and unit $j$, and obeys

$$\begin{aligned}
&\lambda^i_{nj} = 0 \text{ if } \sum_{t=1}^{T} q^{it}_{nj} < g^i_{nj}, \\
&\lambda^i_{nj} > 0 \text{ if } \sum_{t=1}^{T} q^{it}_{nj} = g^i_{nj}.
\end{aligned} \tag{73}$$

The Lagrange multiplier $\lambda^i_{nj}$ represents the *opportunity cost*, i.e., the loss in social welfare incurred by not being able to produce all or a part of $q^{it}_{nj}$ during another interval, which is only relevant when constraint (69) is binding.

Based on the SaT in (39), we can express $\phi^{it}$ by including index $t$ as

$$\begin{aligned}
&\phi^{it}\left(\left[q^{it}_{n'j'}\right]_{n'j'}, \left[q^{i't}_{n'j'}\right]^{i'\neq i}_{n'j'}\right) = U^t\left(\left[q^t_{n'}\right]_{n'}\right) - U^t\left(\left[q^t_{n'} - q^{it}_{n'}\right]_{n'}\right) \\
&- \sum_{n'}\left(P^t_{n'}\left(\left[q^t_{n''}\right]_{n''}\right) q^{it}_{n'} - E_{n'}\left(x^t_{n'}\right) + E_{n'}\left(x^t_{n'} - x^{it}_{n'}\right)\right) + \phi^{it}\left(0, \left[q^{i't}_{n'j'}\right]^{i'\neq i}_{n'j'}\right).
\end{aligned} \tag{74}$$

where $P^t_n\left(\left[q^t_{n'}\right]_{n'}\right)$ is as defined in (14) and indexed by $t$. We can characterise the declared cost function of producer $i$ under the proposed mechanism by reconsidering the spot market generation levels in (13) (indexed by $t$) with the aim of obtaining the solution (72), and obtain

$$\begin{aligned}
\tilde{C}^i_n\left(q^{it}_n\right) = \min_{\left[\bar{q}^{it}_{nj}\right]_j} &\left(\sum_{j'}\left(\hat{C}^i_{nj'}\left(\bar{q}^{it}_{nj'}\right) + \lambda^i_{nj'} q^{it}_{nj'}\right)\right. \\
&+ E_n\left(x^t_n\left(\left[q^{i't}_{nj'}\right]^{i'<i}_{j'}, \left[\bar{q}^{it}_{nj'}\right]_{j'}, \left[q^{i't}_{nj'}\right]^{i'>i}_{j'}\right)\right) \\
&\left. - E_n\left(x^t_n\left(\left[q^{i't}_{nj'}\right]^{i'<i}_{j'}, \left[\bar{q}^{it}_{nj'}\right]_{j'}, \left[q^{i't}_{nj'}\right]^{i'>i}_{j'}\right) - x^{it}_n\left(\left[\bar{q}^i_{nj'}\right]_{j'}\right)\right)\right)
\end{aligned} \tag{75}$$

$$\text{subject to } \bar{q}^{it}_n = q^{it}_n \tag{76}$$

and (71).

Finally, observe that if we set

$$C^i_{nj}\left(q^{it}_{nj}\right) = \hat{C}^i_{nj}\left(q^{it}_{nj}\right) + \lambda^i_{nj} q^{it}_{nj} \tag{77}$$

then we capture the effect of constraint (69) in the cost function. This can be confirmed by observing that the definition of the optimal generation levels in (10) would then be equivalent to (72) and the definition of the declared cost function in (41) to (75). Consequently, the analysis of multiple time intervals under energy limits can be performed for a single interval at a time, as in Sections 3 to 4.2, by using $C^i_{nj}\left(q^{it}_{nj}\right)$ from (77) as the cost function.

Similar to (69), we can introduce additional convex constraints on producer $i$, e.g., a ramping constraint, of the form

$$h\left(\left[q^{it}_{nj}\right]^t_{nj}\right) \leq 0 \tag{78}$$

with an associated Lagrange multiplier $\mu$. Here, $t$ may represent an arbitrary parameter, not necessarily a time interval. Similar to (77), the cost can be offset by $\mu$ as

$$C^i_{nj}\left(q^{it}_{nj}\right) = \hat{C}^i_{nj}\left(q^{it}_{nj}\right) + \mu q^{it}_{nj} \tag{79}$$

As a result, the analysis developed in the previous sections would apply directly and consequently, $\phi^i\left(\left[q^i_{n'j'}\right]_{n'j'}, \left[q^{i'}_{n'j'}\right]^{i'\neq i}_{n'j'}\right)$ from (39) would be incentive compatible.